\documentclass[compsoc, conference, a4paper, 10pt, times]{IEEEtran}

\usepackage{cite}
\usepackage{amsmath,amssymb,amsfonts}
\usepackage{algorithmic}
\usepackage{graphicx}
\usepackage{textcomp}
\usepackage{multirow}
\usepackage{adjustbox}
\usepackage{xcolor}
\usepackage{comment}
\usepackage{booktabs}
\usepackage[inline]{enumitem}
\PassOptionsToPackage{hyphens}{url}
\usepackage[hidelinks]{hyperref}
\usepackage{longtable}
\usepackage{float}
\usepackage{subfiles} % Best loaded last in the preamble

\newif\ifstatus
\statustrue
\newcommand\jesus[1]{\ifstatus\textcolor{black!30!blue}{#1 - Jesus}\fi}

\newcommand\javier[1]{\ifstatus\textcolor{black!30!purple}{#1 - Javier}\fi}

\begin{document}

\title{Threat analysis and adversarial model for Smart Grids}

% Submissions should be anonymized. See the CFP for details on how to anonymize your paper, including any references to your own work.
%\author{\em Anonymous Authors}

% The author information is skipped here, but can be used to include author information in the publication.

\author{\IEEEauthorblockN{Javier Sande-Ríos}
\IEEEauthorblockA{\textit{University Carlos III of Madrid}}
\and
\IEEEauthorblockN{Jesús Canal-Sánchez}
\IEEEauthorblockA{\textit{University Carlos III of Madrid}}
\and
\IEEEauthorblockN{Carmen Manzano-Hernandez}
\IEEEauthorblockA{\textit{University Carlos III of Madrid}}
\and
\IEEEauthorblockN{Sergio Pastrana}
\IEEEauthorblockA{\textit{University Carlos III of Madrid}}
%% IEEE format can accommodate up to six authors this way
}

\maketitle

%%%% PAGE NUMBERING %%%
\thispagestyle{plain}
\pagestyle{plain}

\begin{abstract}
The power grid is a critical infrastructure that allows for the efficient and robust generation, transmission, delivery and consumption of electricity. In the recent years, the physical components have been equipped with computing and network devices, which optimizes the operation and maintenance of the grid. The \textit{cyber} domain of this smart power grid opens a new plethora of threats, which adds to classical threats on the \textit{physical} domain. Accordingly, different stakeholders including regulation bodies, industry and academy, are making increasing efforts to provide security mechanisms to mitigate and reduce cyber-risks. Despite these efforts, there have been various cyberattacks that have affected the smart grid, leading in some cases to catastrophic consequences, showcasing that the industry might not be prepared for attacks from high profile adversaries. At the same time, recent work shows a lack of agreement among grid practitioners and academic experts on the feasibility and consequences of academic-proposed threats. This is in part due to inadequate simulation models which do not evaluate threats based on attackers full capabilities and goals. To address this gap, in this work we first analyze the main attack surfaces of the smart grid, and then conduct a threat analysis from the adversarial model perspective, including different levels of knowledge, goals, motivations and capabilities. To validate the model, we provide real-world examples of the potential capabilities by studying known vulnerabilities in critical components, and then analyzing existing cyber-attacks that have affected the smart grid, either directly or indirectly.
\end{abstract}

% Depending on how vigilant their paper processor is, IEEE may ask for these in your final paper, but we've heard about these amazing inventions called search engines that are able to index every word in your paper, so no need to include them in your submission unless you really want to.

\begin{IEEEkeywords}
Smart Grid, Cybersecurity, Adversarial Model, Power Grid, Critical Infrastructures
\end{IEEEkeywords}

%%%%%%%%%%%%%%%%
% Introduction
%%%%%%%%%%%%%%%%

\section{Introduction}
\label{sec:intro}

%MOTIVATION
The provision of electricity is crucial for the well-being of the society and industry. As such, the power grid is one of the most critical infrastructures for every nation. Yet, recent incidents show that attacks on the Smart Grid are not only possible, but also can cause severe consequences such as black-outs~\cite{mandiant-sandworm-team,threatpost-blackenergy-apt}. Thus, it is essential to understand how attacks might occur, so as to prepare appropriate cyber-physical defenses~\cite{singer2023shedding}. Attacks highly depend on factors such as the attack surface, knowledge, goals and capabilities of an adversary, i.e., the \textit{Adversarial Model}. Understanding these factors allow to better prepare for the potential Techniques, Tactics and Procedures (TTPs) used by cyberattacks, and to foresee how these might evolve in the future. This knowledge, together with the understanding of the organizational critical assets, facilitates the design and implementation of appropriate countermeasures for prevention, detection and risk mitigation~\cite{pastrana2015defidnet}. Still, the literature on Smart Grid security have mostly focused on the analysis of attacks and defenses (see \S\ref{sec:relwork}), without considering the actual motivation, capabilities and knowledge that an adversary might have to conduct such attack. Thus, there is a gap on the analysis of the adversarial model on real world settings. This is one of the reasons that leads to disconnection between real-world, operational and academic-proposed threats, since often academic works rely on incomplete simulated scenarios or unrealistic adversarial models~\cite{singer2023shedding}. 

%CONTRIBUTIONS
To address this gap, in this work we provide a comprehensive analysis of adversaries against the Smart Grid. We first describe the different factors that compose the adversarial model, i.e., attack surface, motivations, goals, knowledge and capabilities. We infer this information from the exiting academic literature, and by studying the history and evolution of cyberattacks on Smart Grids, from which we can analyze the TTPs used, and the goals, targets and impacts of these. We also conduct a study of existing vulnerabilities in critical devices used in operational Smart Grids, which allow us to foresee the potential capabilities that the adversary can exploit to penetrate (and attack) the network. This way, we first formalize an Adversarial Model focusing on the different roles for adversaries against Smart Grids. Then, we use this model to map real-world attacks, e.g., those that targeted and turn-off part of the Ukrainian power grid~\cite{cert-ua-sandworm-attack}.

This paper is structured as follows. First, \S\ref{sec:theoreticalBg} presents a theoretical background about the Smart Grid, and the related work. Then, \S\ref{sec:adverasarialModel} describes the different factors for adversaries, and formalizes an adversarial model. \S\ref{sec:attacks} describes different real-world attacks and maps the proposed adversarial model on these. Finally, \S\ref{sec:conclusions} provides the conclusions of the paper.

%%%%%%%%%%%%%%%%
% Theoretical background
%%%%%%%%%%%%%%%%

\section{Background and Related Work}
\label{sec:theoreticalBg}

This section first describe theoretical concept of the Power Grid and its evolution to the `smart' concept, and then describes the related work.
\subsection{The power grid}
\label{subsec:power-grid}

Electricity is an essential aspect of the society. Consumers have an easy and transparent access to electricity provided by a proper functioning of the electrical grid: a network of synchronized power providers and consumers connected by transmission and distribution lines and operated by one or more control centers. In a nutshell, an electrical grid is composed by: i) \textbf{Power Stations}, for the generation of electrical power (e.g., carbon or nuclear plants, solar panels or wind turbines); ii) \textbf{Electrical Substations}, 
%They may contain one or more generators, which are rotating devices that convert motive or fuel-based power into three-phase electric power. Power stations employ diverse types of energy sources to turn on the generator, being nowadays the most common fossil fuels (e.g., coal, oil, and natural gas), and other options such as nuclear power or renewable energies (e.g., solar panels, wind turbines, or hydroelectric).
that transform high voltage electricity (as generated from the power stations) into low voltage, and vice versa, by means of transformers; iii) \textbf{Electric Power Transmission}, which is the infrastructure, usually transmission lines, that enables the movement of high-voltage electrical energy (i.e., greater than 39kV) from power stations to electric substations; and iv) 
%This usually involve long-distance transmission conducted at high-voltage level (i.e., greater than 39kV) to minimize the loss.%The transmission lines can use either alternating current (AC) or direct current (DC). Usually, AC is employed for shorter transmissions and DC for greater distances as it has fewer losses.
\textbf{Electric Power Distribution}, where the electricity is delivered to the final consumers from a local substation that reduces the high or medium voltage level of the electricity to an usable voltage, typically 230V, which is is then delivered to the final customers.
%The electricity coming from the transmission system goes to a substation, where the high-voltage electricity is transformed to a lower value (between 2 kV and 33 kV), which is the considered range for electric distribution. Once the voltage is reduced in the substation, primary distribution lines carry this medium-voltage electricity to distribution transformers located near the customers. These distribution transformers 
% Since the first central power station was built in Manhattan in 1882, by the Edison Illuminating Company~\cite{pearl-street-station-wikipedia}, \cite{PeakSubstationPowerGrid}, the power grid has evolved in parallel with the engineering and industrial advances during the 20th century. Accordingly, the technological and digital revolution of the last two decades has led to the concept of the \textit{smart} power grid, which is explained next. 

\subsubsection{The smart power grid}
\label{subsec:2-3}

%There is no common definition as to what a Smart Grid is but rather common concepts like their purpose or the advantages it brings, among other things \cite{singer2023shedding}. This section presents an overview of the concept of the Smart Grids.
A \textit{smart} power grid is a cyber-digitally enhanced power grid that 
%uses digital information, communication and control technologies to 
optimize grid operation, leveraging modern Power Automation Systems (PAS), IoT devices and custom communication protocols. 
It allows to reduce cost in the generation, transmission and delivery of electricity, and also enables real-time monitoring of the distribution and demand of electricity~\cite{theta-differences}. 
A key feature introduced by Smart Grids is the decentralization of 
%the bidirectional flow: the electricity is no longer produced only in centralized plants, and then distributed to the consumers. Instead, the 
power production, allowing end-users to become part of the network, e.g., by generating electricity through domestic solar panels. While this reduces costs and energy losses, it simultaneously increases the complexity of the control and management~\cite{gopstein2021nist}. 

%The Smart Grid is an evolving infrastructure, which changes as new technologies emerge. Innovations in the Internet of Things, Artifical Intelligence, or new forms of renewable energy and storage methods will play a key role in the Smart Grid of the future, in order to provide a more sustainable, efficient and cheaper way of managing and distributing electricity.

The Smart Grid involves different entities (i.e., cyber-physical systems, or CPS, computer systems, and the individuals or organizations) operating in different domains~\cite{gopstein2021nist}: the \textbf{Customer} domain is where electricity is mostly consumed, but it can also be produced (e.g., households, commerce, industries, etc.); the \textbf{Market} domain balances the production based on estimated consumption and consumer demands; the \textbf{Service} domain includes tasks such as commercializing, customer management, or installation and maintenance of equipment; the \textbf{Operation} domain, responsible for the smooth function of the Smart Grid, involves tasks such as monitoring, analysis, control, maintenance, etc.;  the \textbf{Generation} domain, where actual production takes place (e.g., coal-fired or nuclear power station), including Distributed Energy Resources (DERs) located at the consumer side (e.g., domestic solar panels); the \textbf{Transmission} domain, focused in the transportation of electricity through different substations; and finally, the \textbf{Distribution} domain, focused on the delivery of electricity from substation to end customers. \S\ref{subsec:Targets} revisits these domains as key attack surfaces for adversaries, and describes the main components exposed on each domain.

\subsection{Related work}
\label{sec:relwork}

The cybersecurity in Smart Grids is an active area of research in academia, with several papers being published each year. As such, there are various surveys on Smart Grid security~\cite{peng2019survey,ding2022cyber,reda2022comprehensive,hasan2023review,nafees2023smart,singer2023shedding}. Since our work is focused on attackers, we review recent works that analyse attacks or adversaries, which we have used to better understand the current landscape and to inform our proposed model.

Peng et al. disscused various attacks at a high granularity (i.e., generation, transmission, distribution or consumption), with general activities (i.e., reconnaissance, scanning, exploitation and access)~\cite{peng2019survey}. Ding J, et al. analyzed vulnerabilities on devices of the Smart Grids from a technical perspective, i.e., the capabilities of the adversary, and also the potential impacts of these, showing real attacks performed on Smart Grids infrastructures worldwide~\cite{ding2022cyber}. Reda et al. focused on existing False Injection Attacks, providing an interesting taxonomy on these attacks, including attack models (e.g., knowledge or capabilities required), targets (e.g., Intelligent Electronic Devices, or IEDs) and impact (i.e., goals)~\cite{reda2022comprehensive}. Kamrul Hasan et al. provide a survey on Smart Grid cybersecurity, including types of attacks that might incur damage on the confidentiality, integrity and availability of data and system on Smart Grids~\cite{hasan2023review}. They also provide a taxonomy of attacks based on the `layer' being affected (i.e., control, communication, physical or cyber layer), and on the impact and goals of these (e.g., economic or physical disruption). Finally, the work of Nafees et al. analyse actual attacks and survey existing countermeasures and to overview existing gaps for cyber-physical situational awareness, including a threat model~\cite{nafees2023smart}. This threat model is composed by an adversary model (i.e., understanding the motivation or resources of different actors), and asset/vulnerability model (i.e., types of components and how these could be exploited), and an attack model (i.e., particularities of attacks, like initial access, propagation or impact). 

Different from previous work, we particularly focus on the adversarial model, considering potential attack surfaces, and the actual capabilities, knowledge, motivation and goals of adversaries. Indeed, a recent paper by Singer et al. shows a disconnection  between real-world operational security (dealing with real systems) and academic works (using simulation and models), showing that threats that could lead to a real impact on the grid are less frequent as those proposed by the academia~\cite{singer2023shedding}. They surveyed cyber-security operators, asking for academic-proposed attacks, and concluded that misperceptions on simulation tools and incomplete models lead to inconsistent scenarios. This work motivated our study, where we propose an adversarial model showing how it can be mapped to real world actors that have attacked the grid (\S\ref{sec:attacks}), and studying vulnerabilities in actual products that could potentially enhance adversarial capabilities (\S\ref{subsec:capabilities}). We believe this is a further step towards more realistic simulations with Smart Grids.

% To the best of our knowledge, no papers that focus on the analysis of vulnerabilities in specific products deployed in the grid have been found. Therefore, to fill this gap, this paper focuses on introducing the study of specific vulnerabilities which, as it will be discussed on 
% \ref{sec:adverasarialModel}, could be of benefit to prevent further attacks on these critical infrastructures. \javier{This is not correct anymore}
% Evolution of cyberattacks

%%%%%%%%%%%%%%%%
% Adversarial model
%%%%%%%%%%%%%%%%

\section{Adversarial model}
\label{sec:adverasarialModel}

When modeling an adversary, it is crucial importance to pinpoint four key elements: their attack surface, objectives, knowledge, and capabilities. We next describe each of these aspects with respect to Smart Power Grids.

\subsection{Attack surface}
\label{subsec:Targets}

The attack surface represent the potential entry points for attacks. As introduced in \S\ref{sec:theoreticalBg}, the Smart Grid is a complex system composed of a diverse range of infrastructures and devices. For the analysis of its characteristics, we will categorize them into the domains defined in \S\ref{subsec:2-3}, where each domain represents a distinct aspect of the Smart Grid infrastructure. Examining these domains individually allows for a focused analysis of specific vulnerabilities and potential attack vectors.

\subsubsection{Power generation}

The generation of energy primarily takes place in power stations. Among these stations, a diverse set of infrastructures exists, involving variations in the type of energy generated (coal, nuclear, wind, solar, etc.), as well as differences in size and age. Many power plants, particularly those dedicated to fossil fuels, exhibit significant age, resulting in original equipment that was not designed with connectivity and cybersecurity in mind. The challenge arises when attempting to adapt and upgrade such outdated systems, posing a formidable obstacle both technically and financially \cite{upgrading-the-power-grid}. The outdated equipment now presents a significant cybersecurity risk, providing malicious actors with opportunities to exploit vulnerabilities. Given the critical role of power generation, in the event of an attack on these infrastructure, an adversarial entity with malicious intentions could manipulate the amount of generated energy, potentially destabilizing the entire network \cite{italian-cyberattack}.

Energy generation can also occur on the consumer side through Distributed Energy Resources (DERs), e.g., solar panels or wind turbines. 

While an attack on a DER would have a smaller impact than an attack on a generation plant, the aggregated effect of a cooperative attack against those generation devices can have a substantial influence on the network~\cite{lindström2021power,dabrowski2017grid}. This poses an escalating risk as the deployment of DERs increases.

\subsubsection{Transmission}

The transmission of electricity, from generation to consumption, involve various systems and infrastructures which are potential targets by adversaries.

\textbf{Substations} are facilities placed along the grid that convert the voltage level of the electricity from high to low or vice versa~\cite{leonardi2014towards}.
%\javier{Removed step-up and step-down explanation. Does it give any crtitical informatión for later explanations or attacks?}
%They can be classified in two types based on their location. First, \textit{step-up} substations, located near the power generation plants, increase the voltage level from low to high for efficient long-distance transmission over high-voltage transmission power lines. Second, \textit{step-down} substations, closer to cities or populated areas, convert the high voltage coming from the transmission power lines into a low voltage suited for its distribution. 
Substations are of various sizes and complexities. Some substations cover small areas and may only contain one bus-bar
%\footnote{A metallic bar in a switchgear panel used to carry electric power.} 
and several circuit breakers. Larger substations cover a significant area and require more components, such as switching, protection and control equipment. Due to the equipment contained and operations performed within, substations are considered a critical part of the power grid and a potential target for attackers~\cite{HUSSAIN2021100406}.

The equipment for this domain primarily consists of \textbf{transformers} and \textbf{protection equipment}. Transformers are used to adapt the voltage levels of electricity to meet the requirements of each grid segment. Meanwhile, protection equipment (e.g., relays, circuit breakers, or disconnect switches) is employed to interrupt the flow of electricity in case of faults or network overload. When these devices detect abnormal operating conditions, they automatically trigger switching equipment to isolate the faulty section and protect the electrical equipment and the grid~\cite{Industroyer2-Video}. %Given the high voltages and exposure to external factors during transmission, 
This equipment plays a critical role in maintaining grid stability and safeguarding physical equipment on the grid or at customer endpoints, and thus, it is an attractive target for attacks aimed at manipulating protection devices, disabling protections, or causing false positives leading to denial of service~\cite{industroyer-v2-mandiant}.

\subsubsection{Operation}

The operation domain include control equipment that serves to monitor and manage the electricity flow, allowing to remotely supervise and maintain the grid's stability. This includes various essential systems. %, like Supervisory Control and Data Acquisition (SCADA) systems or responsible for data collection and supervision from sensors, and Remote Terminal Units (RTUs) enabling remote control over essential equipment of the transmission domain. 
%The use of smart devices in the electric grid allows for remote control and monitoring, leading to a more efficient operation. Nevertheless, it also opens new infection vectors that can be exploited by potential adversaries.
We next describe these systems, including the risks that they are exposed to within the grid and the potential impact than an attack could have on the rest of the structure.

\noindent\textbf{Advanced Metering Infrastructure (AMI)} %facilitates direct communication between end-users and the grid, offering 
provides real-time metrics for energy consumption and production. At its core, AMI relies on smart meters, which receive information about the energy consumption, e.g., from  final users, and other metrics such as battery information, or the amount of energy produced by solar panels~\cite{zheng2013smart}. These devices enable direct communication between end-users and the grid. 
%providing real-time data on energy consumption, as well as energy production form DERs, such as solar panels or energy storage technologies~\cite{khan2022energy}. 
AMI devices are often placed at end-user facilities, and due to ease of physical access, they are highly exposed. Indeed, smart meters are susceptible to physical tampering, leading to fraudulent activities, such as injecting false consumption data to manipulate electricity bills~\cite{badr2023review}. Furthermore, as these meters are often integrated with other systems such supporting apps for remote consumer consultation, they become potential targets for remote attackers~\cite{8926992}.

\noindent\textbf{Geographic Information Systems (GIS)} are responsible for collecting and geographically aggregating real time information from metering devices deployed in the grid~\cite{8671383}. GIS improves decision-making by adding a spatial dimension, helping, for instance, to identify DERs locations or network areas at risks. GIS provides insights which allow the system to react accordingly to maintain the grid stability. GIS establish remote connections with smart meters and other IoT devices, acquiring the necessary data for predictions. However, this exposure poses a risk, as attackers may exploit vulnerabilities in these systems to gather information about connected devices, their topology, or potentially compromise the integrity and confidentiality of received data~\cite{CVE-2021-29114}.

\noindent\textbf{Power Automation Systems (PAS)} are software systems used to monitor electrical substations, retrieving information regarding the part of the grid in which they are deployed. This functionality enables quick and accurate response according to the specific necessities. Moreover, PAS act as an integrator in the grid by incorporating standardized communication protocols, facilitating the exchange of information among different components of the grid. Given their presence across diverse substations within the grid, PAS systems are susceptible to deficiencies in the security measures of these facilities~\cite{7028837,elmasry2023openplc}. Vulnerabilities present on these systems pose a threat to the whole structure of the grid, due to their role in monitoring substation activities and their inherent connection to the devices deployed within it.

\noindent\textbf{Demand Response Systems (DRS)} manage the electricity usage based on supply conditions, pricing, or grid state, leveraging real-time data and communication and optimize energy consumption, costs, and enhancing grid stability~\cite{iea-drs}. The automated response facilitated by smart switches and the integration with AMI, facilitates automatic energy management, ensuring timely responses to grid fluctuations or emergencies~\cite{dabrowski2017grid}.
DRS utilize a combination of hardware devices (such as smart switches) and software solutions responsible for automation and communication with metering and control devices. Consumer devices might include DRS for efficiency (e.g., turning on/off a laundry machine depending on market prices and energy consumption), and thus they are vulnerable to potential compromise, both through the local network of the user or via physical manipulation. This poses a threat to the integrity and confidentiality of the information sent to the Demand Response control infrastructure.

These mentioned systems are composed of both software and hardware equipment for data collection, remote operation and task automation. Key devices integral to this functionality are: 
\begin{enumerate*}[label=\roman*]
\item{\textbf{Supervisory Control And Data Acquisition (SCADA)}} systems, which receive information from different sources and, based on a predefined configuration they manage and act upon the information~\cite{thomas2017power}. 
%They are key within the structure of a Smart Grid, as they receive all the information from substations and control it. Therefore, 
SCADA systems are in the operation center of Smart Grids, thus, an attack on these systems could end in a disruption of the service of the grid. 
%they are vulnerable to compromise on all parts of the grid. 
Additionally, if they are affected by network vulnerabilities, its compromise could lead to other components of the grid being also compromised~\cite{UPADHYAY2020101666};
\item{\textbf{Remote Terminal Unit (RTU)}} are critical components that enable automatic and remote control of grid operations. Moreover, nowadays most of them allow for wireless communication, and thus they are exposed to attackers that can either get in the network through other devices on the same network, or attackers that find vulnerabilities directly in the devices. Since they are a crucial part of the structure of Smart Grids, compromising an RTU would pose at risk the whole network of a Smart Grid~\cite{10.1145/3538969.3544483,dps-rtu};
\item{\textbf{Programmable Logic Controller (PLC)}} are responsible for executing specific tasks based on pre-programmed logic~\cite{8340684}. In Smart Grids PLCs are deployed in various facilities, such as substations and power plants, where they oversee and regulate essential processes. Given their inherent connectivity, they are susceptible to potential cyber threats~\cite{Pickren2024Compromising}. Attackers targeting PLCs could manipulate critical processes, disrupt energy flow, or compromise the integrity of the grid.

\end{enumerate*}

\subsubsection{Market}

The electricity market is highly complex, and its mechanisms vary depending on the region. Nevertheless, in most of these markets, consumption forecasting is a fundamental factor~\cite{6204245}. Since electricity cannot be stored at a large scale, the power grid must maintain a balance between the generation and consumption of electric power. Consumption forecasting allows suppliers to optimize power production, thereby reducing waste and lowering operational costs. Accurate forecasting is also critical in the integration of renewable energy sources, aligning their intermittent generation patterns with the overall demand~\cite{ahmad2020review}.

The diverse consumption patterns of consumers pose a challenge for suppliers, requiring them to align power production with the dynamic nature of real-time consumption. Understanding energy demand is crucial for planning and allocating generation. This prediction relies on algorithms that receive information about past consumption under similar conditions to those being forecasted (same day the previous year, day of the week, weather conditions, etc.)~\cite{6770352}. AMI plays a primary role for the collection of consumption data. 
A failure in forecasting can have serious consequences; overestimation may lead to unnecessary energy generation and economic losses, while underestimation can result in unpreparedness for energy demand and subsequent blackouts. Hence, forecasting algorithms play a pivotal role in the network and may be targeted in attacks aimed at manipulating them~\cite{dabrowski2017grid}.

\subsubsection{Service}

The service domain include user management systems, where sensitive information such as payment data, addresses, and consumption details are stored and managed. Safeguarding this information is crucial to protect user privacy. The systems within this domain share similarities with other ICS, such as water distributors, gas providers, telecommunications companies, etc. and are susceptible to similar types of threats. These attacks may attempt to exploit potential vulnerabilities in their servers to extract customer information, employ social engineering tactics to impersonate a client, or even cause a denial of service, disrupting the network's proper functioning~\cite{bbc-tech-news}. 
%This disruption has the potential to impede payments, overwhelm customer service, or hinder the energy contracting process, as exemplified by the City Power attack~\cite{bbc-tech-news}. The security incident, which will be further analyzed in \S\ref{sec:attacks}, involved a ransomware attack that prevented citizens from purchasing electricity through the pre-paid system, leading to a disruption of the power grid.

\subsubsection{Customer}

The consumer domain is closely related with other domains such as service, market, and operation. Customer consumption and its associated information can be leveraged to manipulate electricity prices and forecast (market), billing statements (service), status and consumption information read by smart meters (operation), or even power generation through DERs. This connection with other domains establishes the consumer as a potential entry point for numerous threats. These include previously mentioned attacks like energy injection through DERs or manipulation of consumption through IoT devices. Furthermore, consumers themselves may become the target of attacks, facing threats such as denial of service, or data leaks that pose risks to their privacy.

A transversal attack surface which affects all the previous domains is the \textbf{human factor} (though differently in terms of the capabilities, goals, and impact that they entail). This includes customers, operators and employees from the Smart Grid. As we describe in \S\ref{sec:attacks}, compound APT attacks often start with a Social Engineering attack targeted at strategically selected employees, i.e., by means of an spearphishing attachment which gives the initial access to the corporate network~\cite{threatpost-blackenergy-apt,nafees2023smart}. 

\subsection{Knowledge}
\label{subsec:knowledge}

Having knowledge of the intricacies of the Smart Grid, including the topology and systems involved, is key to conduct a successful attack. Indeed, the reconnaissance phase is the first step used by adversaries to gain knowledge and plan future attacks~\cite{CISAAdvisoryVoltTyphoon,nafees2023smart}. 
Academic and official resources about the Smart Grid, including international standards, often offer only a high-level description. Moreover, the majority of communication protocols and security measures adhere to public standards, closely aligning with those employed in various industrial systems, albeit with notable complexity. Consequently, a detailed study of these standards and protocols allows an adversary to identify concrete elements and potential exploitation of the communication protocols. 
The electrical grid introduces a degree of opacity at a certain level. While obscurity is not explicitly considered as a security measure, various aspects of the Smart Grid implementation, infrastructure, and topology are deemed confidential by governments~\cite{federal2019critical}. This veil of secrecy surrounding the grid creates a substantial knowledge gap for malicious actors. Consequently attaining a clear understanding of the network's topological structure, operational processes, chain of command in control centers, and their practical implementation remains challenging. 
Based on these considerations, we distinguish three sources of knowledge:

\begin{enumerate}

\item{\textbf{Knowledge through the standards and official sources.}} Adversaries at this level are informed about the established standards that govern the operation and security of the grid~\cite{ISO27019-2017}. Furthermore, attackers can access information about the grid through official documentation publicly available \cite{cmadridTransporte}, which still might be only superficial and lack of details.
Based on this knowledge, adversaries can exploit vulnerabilities and weaknesses of the standards or base their attacks on a estimation of hardware used. Since these standards are open, adversaries could even replicate a production system in order to analyze the system for misconfigurations, weaknesses, or even vulnerabilities in the protocols.

\item{\textbf{Knowledge through the interaction with the grid equipment.}}
The grid is deployed in an open field, and its  infrastructure is accessible at specific locations, such as smart meters in clients' homes, data collectors within residential buildings, or security equipment at the transport domain and substations. This accessibility might be a substantial source for obtaining valuable information about the grid's operation.
Importantly, many of the devices utilized in the Smart Grid are generic and commonly employed in other industrial environments, such as relays, RTUs and SCADA systems. Moreover, detailed documentation for these devices is often publicly available, including their vulnerabilities (see, e.g., \S\ref{subsec:vulnerabilities}). Consequently, the knowledge acquired through interaction with the grid equipment can be combined with publicly accessible information about these devices, empowering potential adversaries to formulate targeted attacks based on a partial understanding of device functionality in the grid.

\item{\textbf{Insider Knowledge}}. Attackers at this level have technical knowledge about both the physical and cyber layers, as well as practical knowledge of the actual implementation of a Smart Grid. Attackers are not only familiar with the standards and the hardware used but also possess a strategic understanding of its topology, organizational structures and emergency response mechanisms, making their potential threats even more sophisticated and challenging to counter.
This depth of knowledge encompasses details about the specific devices utilized on the grid, including their manufacturers and versions, and providing adversaries with a significant advantage in identifying potential attack vectors and vulnerabilities. Such level of insight into the grid is typically obtainable only to individuals with direct involvement in the infrastructure, such as insiders within the targeted grid, or adversaries backed by nations, institutions, or companies capable of providing such intricate details about the power industry.

\end{enumerate}

These knowledge sources are not mutually exclusive. For instance, the understanding of standards can be combined with information gathered from an opportunistic interaction with specific grid devices, providing the adversary with an higher level of knowledge. It is not only the source of knowledge but also its depth that shapes and amplifies the adversary's capabilities to execute an attack.

\subsection{Motivations and Goals}
\label{subsec:goals}

The electrical grid is one of the most valuable critical infrastructure for nations, their institutions and citizens. Also, generation, transportation, and distribution of energy constitute an important economic asset, involving transactions totaling billions of dollars annually in each country. Given the intricate web of stakeholders in the electrical grid, potential cyber threats pose a multifaceted challenge with various objectives and motivations.
Understanding an adversary in this context requires delving into what motivates them to take action, and what specific outcomes they aim to achieve. Thus, we make a distinction between the general motivations for adversaries, and their more targeted and concrete goals.

\subsubsection{Motivations}

We distinguish the following motivations for attacking the grid. 

\noindent\textbf{Geopolitics.} Attackers may seek to exploit the electrical grid as a strategic asset. Motivated by geopolitical considerations, these attackers aim to exert influence, control, or disrupt energy systems to advance broader political goals. The manipulation of the electrical grid can serve as a tool for coercion, influencing regional dynamics, or establishing dominance in the global geopolitical landscape. The electrical grid has become a potent weapon in modern conflicts, employed as a means to weaken adversaries as witnessed in recent conflicts such as the war in Ukraine~\cite{ukraine-invasion}.

\noindent\textbf{Inflict damage on the sector.} Adversaries with this motivation may aim to deliberately sabotage the functioning of the energy sector or a specific company within it.

\noindent\textbf{Inflict harm on the user(s).} Certain actors may target end-users, intending to compromise their safety, privacy, or property through disruptions in their energy supply.

\noindent\textbf{Financial benefit.} For some adversaries, the primary motivation could be financial gains, whether through ransom demands, market manipulations, or other means of economic exploitation.

\noindent\textbf{Fame and recognition.} In some instances, attackers might be driven by the desire for notoriety or acknowledgment within certain circles, seeking recognition for their actions.

\subsubsection{Goals}
Based on their motivations, attackers conduct activities to achieve any of the following goals.

\noindent\textbf{Reconnaissance}. An adversary with this goal seeks to obtain information about critical assets and weakest points of failure. The motivation behind such actions may involve utilizing this information to orchestrate a more intricate attack with a distinct goal or gaining advantages by offering the acquired data to third parties~\cite{nafees2023smart}. This includes information regarding its topology, implemented security measures and critical assets, such as substations, power generation facilities, and control systems. Moreover, the attacker may seek to gather information about human personnel, which can be subsequently employed for social engineering attacks~\cite{blackhat-2022-industroyer2}. Successful reconnaissance enables adversaries to increase their knowledge. 

\noindent\textbf{Service disruption.} This category encompasses attackers aiming to instigate blackouts or disruptions in the electrical grid, potentially causing chaos and impacting various sectors. The objective here could be not only to disrupt daily operations but also to inflict lasting and profound damage, strategically impacting a nation's infrastructure, economy, and overall resilience.

\noindent\textbf{Data theft.} The objective here is to obtain sensitive information, e.g., users' consumption habits. Other valuable data managed by electric companies, such as personal or financial information, may also be targeted. This information can be used to commit other actions, such as blackmailing or fraud, and also it might allow to conduct further steps of a cyberattack.

\noindent\textbf{Market manipulation.} Some adversaries may seek to distort data related to energy consumption, potentially for purposes of influencing market dynamics or cause economic losses.

\noindent\textbf{Manipulating the electricity bill.} A more specific economic goal is to manipulate systems to reduce the cost of bills, e.g., for personal gain or to undermine the provider company. This is often the case for customer fraud.

\subsection{Adversarial capabilities}
\label{subsec:capabilities}

Capabilities are a key factor when characterizing an adversary. They define the spectrum of actions that an attacker can undertake, and in many cases, these coerce the actual goals (i.e., the purpose of the attacks is restricted by the capabilities of the adversary). Also, the capabilities are associated with the knowledge (i.e., the more knowledge and adversary has, the larger capability it has to conduct the attack). Finally, the capabilities highly depend on the attack surface exposed and targeted. 

\subsubsection{Capabilities modeling} 

Following well-known models for capabilities we differentiate between access, exploitation, lateral movement, and persistence~\cite{mitre-attack}. Each category significantly shapes the potential of attackers to achieve their objectives on different phases of the attack.

\subsubsection*{Access} 
The access refers to the attacker's capability to infiltrate the Smart Grid. With \textbf{physical access}, adversaries possess the capability to manipulate or gain entry to targeted devices through their physical layer. This capability may be associated with opportunistic adversaries, such as insiders, or with devices that are not adequately protected at the physical level and exposed to third parties. This might occur either by bad security practices (e.g., an improper access control into a substation), or due to the intrinsic nature of the device  (e.g., a smart meter in end-users' homes, or a transmission cable in the field). The exploitation of physical access requires a knowledge of the physical layer, e.g. to understand how to break in the devices or where to act to conduct a disruption. 
With \textbf{remote access}, attackers possess the capability to infiltrate the intricate networks that constitute the infrastructure. Communication protocols, such as Modbus~\cite{swales1999open}, employed by devices like RTUs or PLCs interfacing with SCADA systems, introduce a diverse range of potential threats. This opens an avenue for virtually any attacker with the required ability to eavesdrop or infiltrate a network, thus jeopardizing the integrity and security of the entire grid. In this case, the adversary requires knowledge of the IT layer and its standards, in order to infer what protocols are being used, and how to exploit weaknesses on these to gain access. Also, recent work propose the use of Web-based PLC malware for modern PLCs, which allows to re-use well-established web-based attacks for industrial PLCs~\cite{Pickren2024Compromising}.

\subsubsection*{Exploitation} The Smart Grid offers multiple exploitation vectors, increasing those for the traditional power grid. These attack vectors are strongly tight to the aforementioned capabilities and most of them are only possible with the corresponding level of access. While acknowledging the vast array of diverse attack strategies and threats, we draw attention to four generic exploitation capabilities:

\begin{itemize}

\item \textbf{Command injection~\cite{10433776}.} With this capability, the adversary might gain control over Smart Grid devices, allowing the execution of malicious commands that can compromise its integrity and operation. While remote command execution poses a risk due to the characteristics of the Smart Grid, this threat also encompasses taking control of devices through physical manipulation (e.g., by an attacker with access to the HMI of a SCADA system, or any mechanical activator of a security device).

\item\textbf{False data injection~\cite{reda2022comprehensive}.} It involves introducing false data into the system, usually to manipulate the recorded consumption. This type of action can have various objectives, such as altering the perception of actual demand, distorting estimations, or even triggering incorrect actions in the operational management of the system leading to a denial of service.

\item\textbf{Denial of service (DoS)~\cite{huseinovic2020survey}.} Attacker can impact the availability of the grid or some of their components. This capability is achieved through various means, which can be related with previously mentioned capabilities, e.g., controlling over devices via command injection, inducing false positives on security devices through the injection of false data, or directly by exploiting vulnerabilities leading to DoS.

\item\textbf{Eavesdropping~\cite{valli2012eavesdropping}.} It is the ability to exploit communication protocols and obtain data of interest, such as user consumption. An attacker with this capability could, for example, intercept data traveling between smart meters and GIS. Also, to observe traffic between control elements and other devices, with the aim of using this information in combination with other capabilities such as false data injection or command injection to, for example, conduct a replay attack.
 Eavesdropping is strongly tied to two primary knowledge sources: standards and direct interaction. First source allows attackers to understand the data being intercepted, whereas second provides insights into how data interception can be achieved. Furthermore, this capability is not solely reliant on knowledge from direct device interaction. It can also be employed to gather additional information about the devices operation and communication mechanisms.

\end{itemize}

\subsubsection*{Lateral Movement and privilege escalation} 
After discussing the potential capabilities of an adversary to access the grid systems and exploit them, we shift our focus to their ability to move laterally to other elements of the network and escalate privileges once they are inside. This capability is crucial in the execution of complex attacks. In attacks on major infrastructures, such as control or generation stations, direct access to control systems from external sources is rarely feasible. Therefore, in most cases attackers gain entry through systems exposed to the public, or through social engineering attacks targeting employees. Once an adversary gains access to one of these devices, it must have the capability to move laterally within internal networks until they successfully access control systems. Leveraging access to local accounts with limited permissions, adversaries exploit systems' vulnerabilities, such as default or hardcoded credentials~\cite{mitre-hardcoded-creds,cve-2016-8566}, gaining access to restricted resources. This capability is closely tied to the adversary's knowledge. The more they know about the internal structure and hierarchy of the organization, the greater their ability to navigate through it. Additionally, operational knowledge of the grid and their communication protocols are essential.

\subsubsection*{Persistence and evasion} 
These capabilities define the attackers' capacity to endure within the compromised systems without detection. This capability varies, ranging from minimal instances where adversaries may execute discrete acts of sabotage, to more sophisticated scenarios where attackers establish clandestine backdoors, providing them with the means to sustain access and control over the system. Once again, we observe a correlation with the attackers' level of knowledge; a deeper understanding of the cyber layer enhances their ability to navigate countermeasures effectively and maintain their activities discreetly. Moreover, possessing detailed knowledge about both layers empowers adversaries to execute Living-of-the-Land attacks~\cite{livingOffTheLand}. In such instances, they leverage existing tools and services within the target environment, evading detection and maintaining persistence~\cite{CISAAdvisoryChinaStateSponsored, CISAAdvisoryVoltTyphoon}.

%\sergio{I think that we should refer further to \url{https://attack.mitre.org/matrices/ics/}, but at the same time, what are the particularities for Smart Grids (oppose to general ICS), giving examples, or referring to the next Section \ref{sec:analysis}}

\subsubsection{Malicious activities}

The extent of damage that can be caused by an attacker employing the mentioned capabilities depends on the nature of the affected devices and the scale of the attack. The malicious activities can span from localized domains, involving the manipulation of user or facility consumption and availability, to broader calamities such as blackouts encompassing city or country-wide areas.
Within the spectrum of potential attacks that can leverage the presented capabilities, we differentiate the following malicious actions:

\noindent\textbf{Modification of Energy Generation.} In instances where an adversary gains remote or physical access to a power generator device, it can manipulate the amount of energy injected into the network, causing an overload or voltage drop~\cite{adepu2020attacks}. These sophisticated attacks require access and control over devices in critical infrastructure, such as generation plants, or the widespread control of DER located at the customer's end~\cite{lindström2021power}. Moreover, the injection of false data could be used to trick the demand-reponse system into responding to a fake generation and consumption imbalance scenario, resulting in the injection of excess or reduced power into the network. Finally, the deactivation of generators can also be achieved through denial of service capabilities targeting the generation plant systems.

\noindent\textbf{Modification of Energy Consumption.} Similar the exploitation of DERs to modify power generation, adversaries can manipulate energy demand to destabilize the network~\cite{dabrowski2017grid}. Seizing control or disrupting the availability of a substantial quantity of loads connected to the grid, such as high-consumption IoT devices, could disturb the delicate balance of the electrical system.

\noindent\textbf{Manipulation of Security Elements.} With the required capabilities an adversary can modify security elements, in order to either cause false positives affecting availability or evade the detection of hazardous situations in the grid. The combination of this manipulation with other attacks, such as altering the amount of energy generated, can not only result in a denial of service but also cause irreversible damage to the infrastructure and connected devices. For example, an adversary with command execution capability over a remote relay can modify the threshold at which current transmission is cut, leading to a blackout in a the area protected by the relay~\cite{8228637}. Similarly, an attacker with false data injection capability could activate a security element by simulating a dangerous state of the grid through manipulated data. Lastly, the availability of security elements can be disrupted by exploiting denial of service capabilities targeted at the vulnerable systems.

\subsubsection{Case study: real-world vulnerabilities} 
\label{subsec:vulnerabilities}

To overview and confirm that the previous capabilities are possible in real world deployments, we conduct a study to gather evidence from 
%the practicality of the described exploiting capabilities over the exposed surface. In doing so, we have taken as a reference
various products, both specific of Smart Grids (e.g., AMI or PAS) and also generic to other ICS (e.g., PLCs and RTUs).
Appendix~\ref{appendix:annex-B} describes the methodology used for this study, related to a selected set of devices from three of the top vendors in this field: Siemens, Esri, and ABB.

Table~\ref{tab:cves} enumerates the number of products considered in devices from these vendors (i.e., GIS, PAS, AMI and DRS), together with the corresponding CPE identifier~\cite{nist-cpe} and CVEs identifier~\cite{mitre-cve}. We provide overall analysis for these vulnerabilities in Appendix~\ref{appendix:annex-B}, and the detailed identifiers in an online appendix.\footnote{\url{https://github.com/jsande-uc3m/adversarial-model-smart-grids}} Here, we provide a classification based on which capabilities would be granted to the adversary, shall the vulnerability be exploited. To achieve this, we categorized them according to their potential for providing access, command and false data injection, eavesdropping, Denial of Service (DoS), lateral movement \& privilege escalation, persistence and other functionalities. This classification is not mutually exclusive, allowing for the potential association of a single vulnerability with multiple capabilities concurrently.

\begin{table}[t]
    \centering
    % \resizebox{\columnwidth}{!}{
    \begin{tabular}{l|c|c|c}
    Component &  \#Products & \#CPEs & \#CVEs \\
    \hline
    Geographic Info. Sys. (GIS) & 16 & 86 & 83 \\ 
    Power Automation Sys. (PAS) & 9 & 15 & 36 \\
    Advance Meter. Infras. (AMI) & 4 + 1* & 14 + 4* & 17 \\
    Demand Resp. Sys. (DRS) & 1 & 4 & 2 \\
    Remote Terminal Unit (RTU) & 7 & 13 & 11\\
    SCADA & 7 & 8 & 40 \\
    Program. Logic Contr. (PLC)  & 5 & 20 & 14 \\
    \hline
    \textbf{Total} & \textbf{50} & \textbf{164} & \textbf{203} \\
    \hline 
    \end{tabular}
    % }
    \caption{Overview CVEs in operation systems and devices}
    \label{tab:cves}
\end{table}

\begin{figure}[t]
    \centering
    \includegraphics[width=.9\columnwidth]{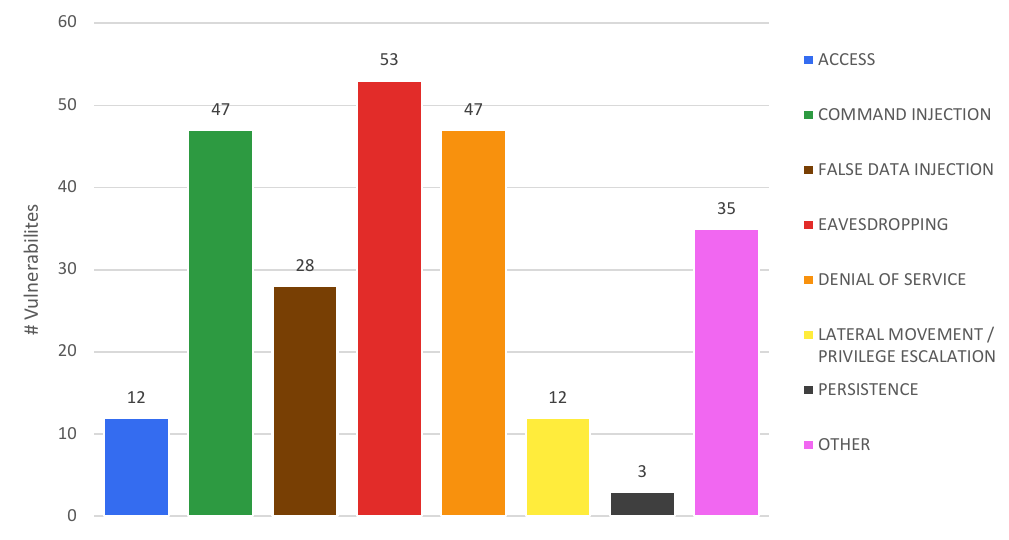}
    \caption{Capabilities enabled by exploiting known vulnerabilities in Smart-Grid systems}
    \label{fig:cve_capabilities}
\end{figure}

The results of this classification are depicted in Figure~\ref{fig:cve_capabilities}. In terms of access, we identified 12 vulnerabilities that, for instance, allow unauthorized users to gain access to resources or administration controls by circumventing authentication systems. Conversely, we also pinpointed a similar number of vulnerabilities facilitating command injection, eavesdropping, and denial of service, with 47, 53, and 47 CVEs respectively. Moreover, we unearthed vulnerabilities breaking data integrity, thereby enabling false data injection. Equally significant as the access vulnerabilities, we uncovered vulnerabilities allowing lateral movement and privilege escalation, empowering authenticated users, for instance, to access resources beyond their permissions or passwords from other accounts. The persistence capability is possible facilitated through 3 vulnerabilities (although it is worth noting that persistence could also be achieved through other capabilities such as command injection). Lastly, categorized as ``others'', we identified 35 vulnerabilities, predominantly linked to social engineering attacks where an adversary redirects legitimate users to rogue websites.

As we can observe, vulnerabilities provide adversaries with the necessary knowledge and potential to exploit security flaws, thereby acquiring the capabilities defined in this section. Some of these vulnerabilities have been exploited in critical  cyberattacks, as we discuss in \S\ref{sec:attacks}.

\subsection{Adversarial model}
\label{subsec:models}

\begin{table*}[t]
\begin{adjustbox}{max width=\textwidth}
\begin{tabular}{|c|c|c|c|c|c|c|c|c|}
\hline
\multirow{2}{*}{\textbf{Role}}& \multirow{2}{*}{\textbf{Motivation}} & \multirow{2}{*}{\textbf{Goals}} & \multirow{2}{*}{\textbf{Knowledge}} & \multicolumn{4}{|c|}{\textbf{Capabilities}}  & \multirow{2}{*}{\textbf{Attack}} \\
\cline{5-8}
& & & & \textbf{Access} & \textbf{Exploitation} 
 & \begin{tabular}[c]{@{}c@{}}\textbf{Lateral}\\ \textbf{movement}\end{tabular} 
 & 
 \begin{tabular}[c]{@{}c@{}}\textbf{Persistence/}\\ \textbf{Evasion}\end{tabular} 
 & \textbf{Surface}\\
  \hline
\textbf{State-sponsored}
& Geopolitics & \begin{tabular}[c]{@{}c@{}}Reconnaissance\\ Service Disruption\\ Data theft\\ Market manipulation\end{tabular} & Insider                                                  & \begin{tabular}[c]{@{}c@{}}Physical\\ Remote\end{tabular} & \begin{tabular}[c]{@{}c@{}}Command Inj.\\ False Data Inj.\\ Eavesdropping\\ DoS\end{tabular} & High                                                       & High                                                               & All domains                                                                             \\ \hline
\textbf{Cyber-terrorist}                                                  & \begin{tabular}[c]{@{}c@{}}Damage sector\\ Harm users\end{tabular} & Service Disruption                                                                                             & \begin{tabular}[c]{@{}c@{}}Official\\ Field\end{tabular} & \begin{tabular}[c]{@{}c@{}}Physical\\ Remote\end{tabular} & \begin{tabular}[c]{@{}c@{}}Command Inj.\\ False Data Inj.\\ Eavesdropping\\ DoS\end{tabular}                 & Medium                                                     & Medium                                                             & Generation                                                                              \\ \hline
\textbf{Cyber-criminal }                                           & Financial benefit                                                  & \begin{tabular}[c]{@{}c@{}}Service Disruption\\ Data theft\\ Market manipulation\end{tabular}                  & \begin{tabular}[c]{@{}c@{}}Official\\ Field\end{tabular} & Remote                                                    & \begin{tabular}[c]{@{}c@{}}Command Inj.\\ DoS\end{tabular}                                   & Medium                                                     & Low                                                                & \begin{tabular}[c]{@{}c@{}}Operation\\ Service\end{tabular}                             \\ \hline
\textbf{Script-kiddie}                                                    & \begin{tabular}[c]{@{}c@{}}Fame and\\ recognition\end{tabular}     & Any                                                                                                            & \begin{tabular}[c]{@{}c@{}}Official\\ Field\end{tabular} & Remote                                                    & \begin{tabular}[c]{@{}c@{}}Eavesdropping\\ DoS\end{tabular}                                  & Low                                                        & Low                                                                & \begin{tabular}[c]{@{}c@{}}Operation\\ Customer\end{tabular}                            \\ \hline
\textbf{Sabotaging insider}
& Damage sector                                                      & Service Disruption                                                                                             & Insider                                                  & \begin{tabular}[c]{@{}c@{}}Physical\\ Remote\end{tabular} & \begin{tabular}[c]{@{}c@{}}Command Inj.\\ False Data In.\\ Eavesdropping\\ DoS\end{tabular}  & Medium                                                     & Medium                                                             & \begin{tabular}[c]{@{}c@{}}Operation\\ Generation\\ Transmission\\ Service\end{tabular} \\ \hline
\textbf{Fraudulent user}
& Financial benefit                                                  & Bill manipulation                                                                                              & Field                                                    & Physical                                                  & \begin{tabular}[c]{@{}c@{}}False Data Inj.\\ DoS\end{tabular}                                & None                                                       & None                                                               & Customer                                                                                \\ \hline
\end{tabular}
\end{adjustbox}
\caption{Smart Grid Adversarial models.\label{long}}
\label{tab:adversarial-models}

\end{table*}

Once we have presented the different features that might define an adversary against Smart Grids, we wrap up with a definition of potential roles based on different models. Table~\ref{tab:adversarial-models} presents and characterize these roles, which we briefly discuss next. %Then, in \S\ref{sec:attacks} we describe real-world attacks and map these to the corresponding adversarial role.

First, \textbf{state-sponsored} refers to the highest profile actor, mostly with geopolitical motivation (i.e, threatening the Smart Grid of an enemy country with any goal). As such, we assume that it has the highest knowledge (i.e., insider, by means of a sabotaging insider or cyber-espionage) and capabilities, and would act on any domain. Then, a \textbf{cyber-terrorist} is an actor that resembles a state-sponsored, but its motivations are more aligned with ideology and its goal is to harm the society or individual users of a targeted victim, i.e., with a main goal of disrupting the generation of energy. Due to potentially having less resources, it might posses fewer capabilities for lateral movement and persistence or evasion. Instances of this threat have been observed~\cite{paganini2015isis}, alongside attacks on the physical layer of energy infrastructure perpetrated by extremists~\cite{usa-extremists}. The next role, \textbf{cyber-criminal}, refers to an actor which seeks to gain financial benefit from the victim (e.g., a ransomware gang), targeting the grid remotely and by means of classical cyberattacks such as command injection or DoS on the operation and service domains. A subset of the previous, less skilled actor, is what we refer as \textbf{script-kiddie}, which different from the previous, it might want to gain  notoriety on underground communities before they evolve towards a higher profile~\cite{pastrana2018characterizing}. An important role, since it might be transversal to the others (e.g., he/she can be coerced or suborned by state-sponsored actors), is the \textbf{sabotage insider}, which mostly refers to an actor with capabilities to threat the sector from the inside, e.g., and employee. Depending on the role and permissions of this actor, its capabilities and targeted domain might differ (e.g., depending on whether he/she has administrative privileges on the industrial network, or if he/she can tamper with particular devices within the grid). Finally, we consider the role of a \textbf{fraudulent user}, which is an adversary that resembles a final customer willing to tamper their local smart meter to report lower readings, and thus to reduce the electricity bill~\cite{badr2023review}.

Overall, our adversarial model helps to understand the different characteristics to understand threats on Smart Grids, and from these, to define different roles. We showcase the benefits of our model by studying real-world attacks, mapping these to the adversarial role.

%%%%%%%%%%%%%%%%
% Attacks
%%%%%%%%%%%%%%%%

\section{Modeling Adversaries from Real World Attacks}
\label{sec:attacks}

This section describes the main cyberattacks that have targeted the power Smart Grid, and maps these attacks to the adversarial model proposed before. To this end, we rely on public information gathered from news and security reports.
%\todo{Explain a bit the methodology followed for this analysis}
%Over the past few years, the power Smart Grid has been the target of many cyberattacks. Some of these attacks have been successfully detected and mitigated on time, causing almost no consequences \javier{Cuantos?}. However, other attacks have resulted in major consequences, such as blackouts or physical damage.
%This section analyses different types of cyberattacks against Power Smart Grids. 
First, we describe targeted malware designed specifically to attack the Smart Grid. Then, we examine more generic malware designed to attack diverse ICS, including ransomware attacks that, despite not being its primary goal, they have impacted the electrical industry with diverse consequences.

\subsection{Cyberattacks targeting the Smart Grid}
\label{subsubsec:3-1-1}
In recent years, the number of attacks against the Smart Grid have increased~\cite{IEA_cybersecurity}. 
Table \ref{tab:malware-attacks-1} presents a list of the major cyber-incidents that have directly targeted and caused significant impact on power grids. As it can be observed, they all targeted the power grid of Ukraine.
The BlackEnergy-3 malware used in the 2015 Ukrainian blackout is not a malware specifically designed to attack power grids but is a multi-purpose malware that has been active since 2007 (in earlier versions) and was used since then in multiple campaigns. In contrast, the Industroyer and Industroyer-2 malware pieces, used in the 2016 and 2022 attacks respectively, were specifically designed to attack power grids. This means that they are a more sophisticated and dangerous threat to power grids because they attempt to exploit specific features and protocols used in electrical substations. 
These attacks, summarized in Table \ref{tab:malware-attacks-1}, have the same pattern, i.e., they all have geopolitical motivation and the goal for disruption. Despite subtle differences, mostly in terms of sophistication, they all fall under the same adversarial model, i.e., state-sponsored. We next describe the first attack (2015), mapping each of the actions to the proposed adversarial model.
%\sergio{Consider adding an extended version to arXiv and cite this version}
%, an in-depth analysis will be carried out in Section \ref{subsec:4-1}. \sergio{I brought back the specific analyses of these 3 attacks. We need to reduce the description and map the attacks better to the adversarial model }

\begin{table}
\centering
\begin{tabular}{lp{2.5cm}p{1.8cm}l}
\toprule
\textbf{Year} & \textbf{Target} & \textbf{Cyber-weapon} & \textbf{Impact} \\
\midrule
2015 & Various substations (Ukraine) & BlackEnergy-3  & Blackout \\
2016 & Pivnichna substation (Ukraine) & Industroyer / CrashOverride  & Blackout \\
2022 & Unknown (Ukraine) & Industroyer-2 & None\textsuperscript{\textdagger} \\
\bottomrule
\end{tabular}
\caption{Cyberattacks targeting the Ukranian Smart Grid for service disruption. (\textsuperscript{\textdagger}The attack was stopped by the national CERT before incurring any further damage)}
\label{tab:malware-attacks-1}
\end{table}

%\jesus{He quitado contenido y reorganizado partes de esta sección para reducir su tamaño}
%\label{subsec:4-1}

\subsubsection{Ukrainian blackout (2015)}
\label{subsubsec:4-1-1}
The first known cyberattack that caused a blackout in a populated city took place in Ukraine on December 23, 2015. The cyberattack caused several power outages, which affected 225,000 customers approximately, in different regions of Ukraine, and lasted for almost 6 hours~\cite{case2016analysis} \textbf{[goal-disrupt]}. The attack was attributed by various security firms and hacking experts to the Russian hacking group Sandworm~\cite{mandiant-sandworm-team} \textbf{[motivation-geopolitical}].
%Due to the malware employed in this %attack, it is also believed that the Sandworm group collaborated with the BlackEnergy APT %group to carry out the attack \cite{threatpost-blackenergy-apt}.

This attack that led to the black-out was not an isolated incident, but rather a continuation of multiple attacks carried out by Russia against Ukraine's critical infrastructure. Indeed, the first step the attackers took before executing the actual attack was a reconnaissance phase \textbf{[goal-reconnaissance]}, in which they tried to gather all possible information about the potential targets \textbf{[goal-data-theft]}. Once they had all the necessary information, the actors developed custom malware for the attack. The adversaries employed spear-phishing against employees of the different Ukrainian substations \textbf{[capability-false-data-injection]}, sending e-mails with weaponized Microsoft Office documents with an embedded installer of the BlackEnergy3 (BE-3) malware~\cite{styczynski2019lights}. 
%This installer was triggered by enabling the macros prompt, which allowed its execution. The experts that analyzed the malware samples from the attack also found evidence of the use of the Killdisk malware to destroy data and hardware \cite{styczynski2019lights}. 
%In this case, as mentioned before, the attackers selected the BE-3 malware and chose weaponized MS Office documents as the form of distribution. 
The BE-3 malware established connection to its Command and control (C2C) server for further instructions \textbf{[capability-command-injection]}. After reaching this point, they performed lateral movement through the internal corporate network in order to discover new targets and expand the invasion \textbf{[capability-lateral]}, and they also gained access to critical devices within the ICS network. 
Allegedly, the attackers reached this point around six months before the actual date of the black-out, collecting necessary data and gaining knowledge from the infrastructure to conduct the attack (\textbf{[knowledge-insider]}).

Once this phase was reached, the hackers placed the KillDisk malware on a network share and added a policy in the domain controller to retrieve this malware and execute upon system reboots \textbf{[capability-persistence]}. Next, they prepared a new attack to launch in parallel to the actual black-out. This attack consisted on cutting the uninterruptible power supply (UPS) of the telephone communications server and the substation data center servers to prevent users from reporting the loss of power and thus making the outage last longer \textbf{[capability-DoS]}.

For the attack, the attackers ``pulled the plug'' by remotely controlling the computers of the substation’s employees and issuing unauthorized commands \textbf{[capability-command-injection]} that opened the circuit breakers, interrupting the power supply to more than 225,000 customers \textbf{[goal-disrupt]}. Once the hackers cut the power, they delivered a malicious firmware update to the target devices to disable serial to Ethernet converters. Consequently, operators were unable to remotely close the breakers, requiring to manually close the breakers at each substation, increasing the recovery time and the impact of the attack~\textbf{[capability-persistence],[capability-DoS]}. 
Also, the attackers cut the phone communication server and data center server of the substations through a DoS attack, tampering the reporting for the customers\textbf{[capability-evasion],[capability-DoS]}. Once the attack took place, the actors executed the KillDisk malware again to erase all the records and log data from the victim’s machines from both the corporate and ICS networks~\cite{case2016analysis} \cite{styczynski2019lights} \textbf{[capability-evasion],[capability-DoS]}.

\subsection{Cyberattacks affecting the power grid}
\label{subsubsec:3-1-2}
In this section, we describe cyberattacks that had an impact on the power grid, but with a lower level of planning and sophistication than the targeted cyberattacks described previously. Table \ref{tab:malware-attacks-2} shows the attacks that employ more-generic malware, not explicitly designed to attack power Smart Grids.  We next discuss them in detail.

In 2016 a cyberattack against the Israel Electric Authority was reported by the national government, claiming that they were targeted by severe cyberattacks causing several computers to shut down. In response, officials chose to switch off segments of the country's power grid, hindering its operation~\cite{TrendMicroElectricAuthorityHack} \textbf{[goal-disruption]}.
A later report from Dragos Security contradicts the version given by the government, claiming that the cyberattack was actually an undisclosed ransomware delivered via spearphishing that infected an employee from Israel’s Electric, taking off some computers from the Electric Authority but not incurring into important outages affecting the power grid~\textbf{[goal-financial],[capability-dos]}. Since no further information was disclosed, we can not model the potential adversary. However, this case shows how a cyber-criminal, potentially seeking financial gain, was able to alarm an energy operator, ultimately leading to a deprivation of a national critical infrastructure.
%This case exemplifies important operational security flaws and erroneous or inaccurate %official communications from the governments.

The 2017 attack against EirGrid, a power company that provides electricity across Ireland and Northern Ireland, resembles a state-sponsored attack, though there is not official attribution. According to multiple reports, adversaries gained access by using a MITM (Man-In-The-Middle) technique (\textbf{[capability-eavesdropping]}) and installing a virtual tap on Vodafone's Direct Internet Access (DIA) service in Shotton, Wales~\cite{bitdefender-blog}. The wiretaps were discovered in July 2017 and have allegedly been active since April of that same year. In these 4 months, hackers were able to intercept communications and steal information (\textbf{[goal-data-theft],[knowledge-insider]}. The extent of data compromised by the hackers and the potential installation of additional malware remain undisclosed. However, as mentioned previously, attackers might seek to gain knowledge and persist before conducting further attacks \textbf{[goal-reconnaisance]}. At the time of this writing (March'24) no further information have been made public regarding these incidents in Ireland's power grid \cite{bitdefender-blog}. 

In 2019, a ransomware attack against City Power, a major electricity provider in South Africa, encrypted all their databases, applications and network \textbf{[motivation-financial],[capability-DoS]}. The attack disrupted prepaid customers' ability to purchase electricity units, consequently leading to an eventual shortage of supply \textbf{[goal-disruption]}. The number of customers who were affected by this problem was over 250,000. The ransomware used, whereas it has been allegedly identified, has not been disclosed publicly at the time of this writing \cite{bbc-tech-news}. This is another example of a cybercriminal adversary, that, seeking for financial benefit, and by targeting the service domain, has disrupted a critical infrastructure due to the way the market domain was designed. 

\begin{table}
\centering
\begin{tabular}{lp{3cm}ll}
\toprule
\textbf{Year} & \textbf{Target} & \textbf{Cyber-weapon} & \textbf{Impact}\\
\midrule
2016 & Electricity Authority (Israel) & Ransomware & Operability\\

2017 & EirGrid (Ireland) & None (MITM) & Unknown\textsuperscript{\textdagger} \\

2019 & City Power (South Africa) & Ransomware & Operability\\
\bottomrule
\end{tabular}
\caption{Cyberattacks with negative effects on the electric operation. (\textsuperscript{\textdagger}Allegedly, data leakage or malware installation)}
\label{tab:malware-attacks-2}
\end{table}

\subsubsection{Ransomware attacks that hit the electrical sector}
\label{subsubsec:3-1-5}
Various cyberattacks, even not directly targeting the grid operation (i.e., distribution, generation or transmission domains), have affected the electricity industry, concretely the service, customer and market domains.
% Due to high volume of ransomware attacks affecting companies from the electricity sector, it is out the scope of our paper to conduct an analysis of each particular case, except for those mentioned in Section \ref{subsubsec:3-1-2} (as in those cases, the consequences they caused were more severe than financial or minor disruptions).
Table \ref{tab:ransomware-attacks} in Appendix \ref{appendix:annex-A} includes  the ransomware attacks against utilities in the electricity sector that could be identified. For this analysis, we rely on the CIRA dataset~\cite{rege2023critical}. Even though the dataset starts taking data from 2017 onward, the number of ransomware attacks against companies in the electrical sector has been increasing since 2017. This is expected and consistent with the popularity and prevalence of these attacks in the last years~\cite{smh-gru-explained}.
%\sergio{CITE NEEDED}. Indeed, according to Dragos' 2022 annual report, there has been a 30\% increase over the previous year in the number of active ransomware groups and an 87\% increase in ransomware attacks targeting industrial organizations, to a great extent due to political and financial motivation, and also the continued growth of the RaaS (Ransomware as a Service) industry~\cite{smh-gru-explained}.
The most common approach used for initial exploitation is via phishing e-mails to company employees \textbf{[capability-false-data-injection]}. The consequences for companies are, in most cases, financial losses, e.g., payments to recover the data or prevent data leakage, and also reputational, since these attacks degrade the confidence and trust into these companies \textbf{[goal-financial],[goal-market-manipulation]}. However, in some minor but relevant cases, such as the ones discussed before, these attacks entail operational problems, meaning that their cyber-physical systems have been temporarily shut down because of the attack.

\section{Conclusions and discussion}
\label{sec:conclusions}

The Smart Grid is a critical infraestructure that needs to be properly protected, and the attacks in the Ukrainian grid showed the impact that a high-profile adversary can incur on the society. To better prepare, detect and respond to these incidents, it is important to understand the attacks surfaces, as well a the knowledge, capabilities, motivations and goals of the adversary, i.e., to know the adversarial model. We define a model by considering attacks from the research literature, exiting vulnerabilities, and by analysing the Tactics, Techniques and Procedures (TTPs) from real-world attacks. This model allows to define different actors (roles) for these incidents, with different levels of skills and goals. Our work allows to better understand the risks to which the Smart Grid is exposed, and consequently, to apply appropriate and reasonable countermeasures to mitigate these.

%%%%%%%%%%%%%%%%
% Acknowledgments
%%%%%%%%%%%%%%%%

 \section*{Acknowledgments}
As part of the open-report model followed by the Workshop on Attackers \& CyberCrime Operations (WACCO), all the reviews for this paper are publicly available at \url{https://github.com/wacco-workshop/WACCO/tree/main/WACCO-2024}. This work was supported by Grant TED2021-132170A-I00 funded by MCIN/AEI/ 10.13039/501100011033 and by the ``EU NextGenerationEU/PRTR''.

%%%%%%%%%%%%%%%%
% Bibliography
%%%%%%%%%%%%%%%%

\bibliographystyle{plain}
\bibliography{references}

\begin{thebibliography}{10}

\bibitem{adepu2020attacks}
Sridhar Adepu, Nandha~Kumar Kandasamy, Jianying Zhou, and Aditya Mathur.
\newblock Attacks on smart grid: Power supply interruption and malicious power generation.
\newblock {\em International Journal of Information Security}, 19:189--211, 2020.

\bibitem{ahmad2020review}
Tanveer Ahmad, Hongcai Zhang, and Biao Yan.
\newblock A review on renewable energy and electricity requirement forecasting models for smart grid and buildings.
\newblock {\em Sustainable Cities and Society}, 55:102052, 2020.

\bibitem{badr2023review}
Mahmoud~M Badr, Mohamed~I Ibrahem, Hisham~A Kholidy, Mostafa~M Fouda, and Muhammad Ismail.
\newblock Review of the data-driven methods for electricity fraud detection in smart metering systems.
\newblock {\em Energies}, 16(6):2852, 2023.

\bibitem{bbc-tech-news}
{BBC News}.
\newblock Ransomware hits johannesburg electricity supply, 2019.
\newblock Accessed: 30-05-2023.

\bibitem{threatpost-blackenergy-apt}
Chris Brook.
\newblock Blackenergy apt group spreading malware via tainted word docs, 2016.
\newblock Accessed: 16-03-2023.

\bibitem{italian-cyberattack}
Clementina Bruno, Luca Guidi, Azahara Lorite-Espejo, and Daniela Pestonesi.
\newblock Assessing a potential cyberattack on the italian electric system.
\newblock {\em IEEE Security \& Privacy}, 13(5):42--51, 2015.

\bibitem{cert-ua-sandworm-attack}
{CERT-UA}.
\newblock Cyberattack of the sandworm group (uac-0082) on energy facilities of ukraine using industroyer2 and caddywiper malware (cert-ua-4435), 2022.
\newblock Accessed: 04-03-2023.

\bibitem{blackhat-2022-industroyer2}
Anton Cherepanov and Robert Lipovsky.
\newblock Industroyer 2 - sandworm's cyberwarfare targets ukraine's power grid again, 2022.
\newblock Accessed: 04-03-2023.

\bibitem{Industroyer2-Video}
Robert Lipovsky \&~Anton Cherepanov.
\newblock Industroyer2: Sandworm's cyberwarfare targets ukraine's power grid again, 2022.
\newblock Accessed: 13-03-2023.

\bibitem{bitdefender-blog}
Graham Clueley.
\newblock Attack on ireland's state-owned power provider blamed on state-sponsored hackers, 2017.
\newblock Accessed: 30-05-2023.

\bibitem{federal2019critical}
Federal Energy~Regulatory Commission et~al.
\newblock Critical energy/electric infrastructure information (ceii).
\newblock \url{https://www.ferc.gov/ceii}, 2019.
\newblock Accessed: 2024-02-22.

\bibitem{mitre-attack}
Mitre Corporation.
\newblock Mitre att\&ck.
\newblock \url{https://attack.mitre.org/}, -.
\newblock Accessed: 2024-02-16.

\bibitem{mitre-hardcoded-creds}
Mitre Corporation.
\newblock Hardcoded credentials.
\newblock \url{https://attack.mitre.org/techniques/T0891/}, 2022.
\newblock Accessed: 2024-03-16.

\bibitem{mitre-cve}
The~MITRE Corporation.
\newblock Common vulnerabilities and exposures.
\newblock \url{https://www.cve.org}, -.
\newblock Accessed: 2024-03-16.

\bibitem{dabrowski2017grid}
Adrian Dabrowski, Johanna Ullrich, and Edgar~R Weippl.
\newblock Grid shock: Coordinated load-changing attacks on power grids: The non-smart power grid is vulnerable to cyber attacks as well.
\newblock In {\em Proceedings of the 33rd Annual Computer Security Applications Conference}, pages 303--314, 2017.

\bibitem{cmadridTransporte}
Comunidad de~MADRID.
\newblock Planificación de la red de transporte de electricidad 2015-2020.
\newblock \url{https://www.ree.es/sites/default/files/01_ACTIVIDADES/Documentos/planificacion/madrid_v2.pdf}, 2014.
\newblock Accessed: 2024-02-22.

\bibitem{ding2022cyber}
Jianguo Ding, Attia Qammar, Zhimin Zhang, Ahmad Karim, and Huansheng Ning.
\newblock Cyber threats to smart grids: Review, taxonomy, potential solutions, and future directions.
\newblock {\em Energies}, 15(18), 2022.

\bibitem{case2016analysis}
E-ISAC.
\newblock Analysis of the cyber attack on the ukrainian power grid, 2016.
\newblock Accessed: 14-02-2023.

\bibitem{elmasry2023openplc}
Ahmed Elmasry, Abdullatif Albaseer, and Mohamed Abdallah.
\newblock Openplc and lib61850 smart grid testbed: Performance evaluation and analysis of goose communication.
\newblock In {\em 2023 International Symposium on Networks, Computers and Communications (ISNCC)}, pages 1--6. IEEE, 2023.

\bibitem{smh-gru-explained}
Guy Faulconbridge.
\newblock Russia's gru military intelligence agency explained, 2018.
\newblock Accessed: 08-03-2023.

\bibitem{usa-extremists}
Eric Geller.
\newblock Physical attacks on electrical grid peak amid cyber threats.
\newblock {\em Politico}, December 2022.

\bibitem{8671383}
Lavanya Gnanasekaran and Sean Monemi.
\newblock Gis role in smart grid.
\newblock {\em 2018 IEEE Conference on Technologies for Sustainability (SusTech)}, pages 1--5, 2018.

\bibitem{gopstein2021nist}
Avi Gopstein, Cuong Nguyen, Cheyney O'Fallon, Nelson Hastings, David Wollman, et~al.
\newblock {\em NIST framework and roadmap for smart grid interoperability standards, release 4.0}.
\newblock Department of Commerce. National Institute of Standards and Technology~…, 2021.

\bibitem{8926992}
Yutian Gui, Ali~Shuja Siddiqui, Suyash~Mohan Tamore, and Fareena Saqib.
\newblock Security vulnerabilities of smart meters in smart grid.
\newblock In {\em IECON 2019 - 45th Annual Conference of the IEEE Industrial Electronics Society}, volume~1, pages 3018--3023, 2019.

\bibitem{hasan2023review}
Mohammad~Kamrul Hasan, AKM~Ahasan Habib, Zarina Shukur, Fazil Ibrahim, Shayla Islam, and Md~Abdur Razzaque.
\newblock Review on cyber-physical and cyber-security system in smart grid: Standards, protocols, constraints, and recommendations.
\newblock {\em Journal of Network and Computer Applications}, 209:103540, 2023.

\bibitem{huseinovic2020survey}
Alvin Huseinovi{\'c}, Sa{\v{s}}a Mrdovi{\'c}, Kemal Bicakci, and Suleyman Uludag.
\newblock A survey of denial-of-service attacks and solutions in the smart grid.
\newblock {\em IEEE Access}, 8:177447--177470, 2020.

\bibitem{HUSSAIN2021100406}
Shahbaz Hussain, Javier {Hernandez Fernandez}, Abdulla~Khalid Al-Ali, and Abdullatif Shikfa.
\newblock Vulnerabilities and countermeasures in electrical substations.
\newblock {\em International Journal of Critical Infrastructure Protection}, 33:100406, 2021.

\bibitem{iea-drs}
iea.
\newblock Demand response.
\newblock \url{https://www.iea.org/energy-system/energy-efficiency-and-demand/demand-response}, 2023.
\newblock Accessed: 2023-10-09.

\bibitem{IEA_cybersecurity}
{International Energy Agency}.
\newblock Cybersecurity: is the power system lagging behind?, 2023.

\bibitem{ISO27019-2017}
{International Organization for Standardization (ISO)}.
\newblock Information technology -- security techniques -- information security management guidelines based on iso/iec 27002 for process control systems specific to the energy utility industry, 2017.

\bibitem{energydigital-research-companies}
Amber Jackson.
\newblock Top 10: Smart grid companies.
\newblock \url{https://energydigital.com/top10/top-10-smart-grid-companies}, 2023.
\newblock Accessed: 2023-11-09.

\bibitem{leonardi2014towards}
A~Leonardi, K~Mathioudakis, A~Wiesmaier, and F~Zeiger.
\newblock Towards the smart grid: substation automation architecture and technologies.
\newblock {\em Advances in electrical engineering}, 2014, 2014.

\bibitem{lindström2021power}
Martin Lindström, Hampei Sasahara, Xingkang He, Henrik Sandberg, and Karl~Henrik Johansson.
\newblock Power injection attacks in smart distribution grids with photovoltaics, 2021.

\bibitem{mandiant-sandworm-team}
{Mandiant}.
\newblock Sandworm team and the ukrainian power authority attacks, 2016.
\newblock Accessed: 11-03-2023.

\bibitem{6770352}
Piotr Mirowski, Sining Chen, Tin~Kam Ho, and Chun-Nam Yu.
\newblock Demand forecasting in smart grids, 2014.

\bibitem{6204245}
Amir Motamedi, Hamidreza Zareipour, and William~D. Rosehart.
\newblock Electricity price and demand forecasting in smart grids.
\newblock {\em IEEE Transactions on Smart Grid}, 3(2):664--674, 2012.

\bibitem{nafees2023smart}
Muhammad~Nouman Nafees, Neetesh Saxena, Alvaro Cardenas, Santiago Grijalva, and Pete Burnap.
\newblock Smart grid cyber-physical situational awareness of complex operational technology attacks: A review.
\newblock {\em ACM Computing Surveys}, 55(10):1--36, 2023.

\bibitem{nist-cpe}
NIST.
\newblock Official common platform enumeration (cpe) dictionary.
\newblock \url{https://nvd.nist.gov/products/cpe}, -.
\newblock Accessed: 2023-10-09.

\bibitem{cve-2016-8566}
NIST.
\newblock Cve-2016-8566 detail, 2017.
\newblock Accessed: 2024-03-16.

\bibitem{CVE-2021-29114}
NIST.
\newblock Cve-2021-29114.
\newblock \url{https://nvd.nist.gov/vuln/detail/CVE-2021-29114}, 2021.
\newblock Accessed: 2024-02-22.

\bibitem{nist-score-severity}
NIST.
\newblock Vulnerability metrics, 2022.
\newblock Accessed: 2023-10-09.

\bibitem{ukraine-invasion}
Patrick~Howell O'Neill.
\newblock Russian hackers tried to bring down ukraine’s power grid to help the invasion.
\newblock \url{https://www.technologyreview.com/2022/04/12/1049586/russian-hackers-tried-to-bring-down-ukraines-power-grid-to-help-the-invasion/}, 2022.
\newblock Accessed: 2024-03-16.

\bibitem{paganini2015isis}
Pierluigi Paganini.
\newblock Isis cyber caliphate.
\newblock {\em Security Affairs}, 2015.

\bibitem{pastrana2018characterizing}
Sergio Pastrana, Alice Hutchings, Andrew Caines, and Paula Buttery.
\newblock Characterizing eve: Analysing cybercrime actors in a large underground forum.
\newblock In {\em Research in Attacks, Intrusions, and Defenses: 21st International Symposium, RAID 2018, Heraklion, Crete, Greece, September 10-12, 2018, Proceedings 21}, pages 207--227. Springer, 2018.

\bibitem{pastrana2015defidnet}
Sergio Pastrana, Juan~E Tapiador, Agustin Orfila, and Pedro Peris-Lopez.
\newblock Defidnet: A framework for optimal allocation of cyberdefenses in intrusion detection networks.
\newblock {\em Computer Networks}, 80:66--88, 2015.

\bibitem{peng2019survey}
Chen Peng, Hongtao Sun, Mingjin Yang, and Yu-Long Wang.
\newblock A survey on security communication and control for smart grids under malicious cyber attacks.
\newblock {\em IEEE Transactions on Systems, Man, and Cybernetics: Systems}, 49(8):1554--1569, 2019.

\bibitem{Pickren2024Compromising}
Ryan Pickren, Tohid Shekari, Saman Zonouz, and Raheem Beyah.
\newblock Hey, you, get off of my market: detecting malicious apps in official and alternative android markets.
\newblock In {\em NDSS}, 2024.

\bibitem{theta-differences}
Theta~Learning Point.
\newblock Difference between smart grid and conventional grid.
\newblock \url{https://www.thetalearningpoint.com/2023/01/difference-between-smart-grid-and-conventional-grid.html}, 2023.
\newblock Accessed: 2023-10-09.

\bibitem{reda2022comprehensive}
Haftu~Tasew Reda, Adnan Anwar, and Abdun Mahmood.
\newblock Comprehensive survey and taxonomies of false data injection attacks in smart grids: attack models, targets, and impacts.
\newblock {\em Renewable and Sustainable Energy Reviews}, 163:112423, 2022.

\bibitem{rege2023critical}
A.~Rege.
\newblock Critical infrastructure ransomware attacks (cira) dataset.
\newblock Online, 2023.
\newblock Accessed: 26-04-2023.

\bibitem{10.1145/3538969.3544483}
Engla Rencelj~Ling, Jose~Eduardo Urrea~Cabus, Ismail Butun, Robert Lagerstr\"{o}m, and Johannes Olegard.
\newblock Securing communication and identifying threats in rtus: A vulnerability analysis.
\newblock In {\em Proceedings of the 17th International Conference on Availability, Reliability and Security}, ARES '22, New York, NY, USA, 2022. Association for Computing Machinery.

\bibitem{emergen-research-companies}
Emergen Research.
\newblock Top 10 companies advancing smart grid technology to create sustainable energy future.
\newblock \url{https://www.emergenresearch.com/blog/top-10-companies-advancing-smart-grid-technology-to-create-sustainable-energy-future}, 2023.
\newblock Accessed: 2023-11-09.

\bibitem{upgrading-the-power-grid}
Carla Rubí.
\newblock The challenges of upgrading the power grid for a decarbonised electric future.
\newblock \url{https://informaconnect.com/the-challenges-of-upgrading-the-power-grid-for-a-decarbonised-electric-future/}, 2019.
\newblock Accessed: 2024-02-22.

\bibitem{8340684}
Gabriel Salles-Loustau, Luis Garcia, Pengfei Sun, Maryam Dehnavi, and Saman Zonouz.
\newblock Power grid safety control via fine-grained multi-persona programmable logic controllers.
\newblock In {\em 2017 IEEE International Conference on Smart Grid Communications (SmartGridComm)}, pages 283--288, 2017.

\bibitem{siemens-catalog}
Siemens.
\newblock Energy automation – intelligent and future-proof.
\newblock \url{https://www.siemens.com/global/en/products/energy/energy-automation-and-smart-grid.html}, -.
\newblock Accessed: 2023-09-09.

\bibitem{siemens-esri}
Siemens.
\newblock Siemens and esri partner to bring grid planning and operation to a new level.
\newblock \url{https://press.siemens.com/global/en/pressrelease/siemens-and-esri-partner-bring-grid-planning-and-operation-new-level}, 2022.
\newblock Accessed: 2023-09-09.

\bibitem{singer2023shedding}
Brian Singer, Amritanshu Pandey, Shimiao Li, Lujo Bauer, Craig Miller, Lawrence Pileggi, and Vyas Sekar.
\newblock Shedding light on inconsistencies in grid cybersecurity: Disconnects and recommendations.
\newblock In {\em 2023 IEEE Symposium on Security and Privacy (SP)}, pages 38--55. IEEE, 2023.

\bibitem{spring2021time}
Jonathan Spring, Eric Hatleback, Allen Householder, Art Manion, and Deana Shick.
\newblock Time to change the cvss?
\newblock {\em IEEE Security \& Privacy}, 19(2):74--78, 2021.

\bibitem{styczynski2019lights}
Jake Styczynski and Nate Beach-Westmoreland (Booz~Allen Hamilton).
\newblock When the lights went out. a comprehensive review of the 2015 attacks on ukrainian critical infrastructure.
\newblock {\em Booz Allen Hamilton}, 2019.
\newblock Accessed: 15-03-2023.

\bibitem{swales1999open}
Andy Swales et~al.
\newblock Open modbus/tcp specification.
\newblock {\em Schneider Electric}, 29(3):19, 1999.

\bibitem{livingOffTheLand}
{Symantec}.
\newblock Living off the land: Turning your infrastructure against you.
\newblock Online: \url{https://www.symantec.com/content/dam/symantec/docs/white-papers/living-off-the-land-turning-your-infrastructure-against-you-en.pdf}, 2019.
\newblock Accessed: 16-03-2024.

\bibitem{dps-rtu}
DPS Telecom.
\newblock A rugged remote terminal unit for the smart grid market.
\newblock \url{https://www.dpstele.com/insights/2020/01/03/smart-grid-market/}, 2020.
\newblock Accessed: 2023-10-09.

\bibitem{thomas2017power}
Mini~S Thomas and John~Douglas McDonald.
\newblock {\em Power system SCADA and smart grids}.
\newblock CRC press, 2017.

\bibitem{TrendMicroElectricAuthorityHack}
{Trend Micro}.
\newblock Israel's electric authority "hack" caused by ransomware, 2016.
\newblock Accessed: 29-05-2023.

\bibitem{UPADHYAY2020101666}
Darshana Upadhyay and Srinivas Sampalli.
\newblock Scada (supervisory control and data acquisition) systems: Vulnerability assessment and security recommendations.
\newblock {\em Computers \& Security}, 89:101666, 2020.

\bibitem{CISAAdvisoryChinaStateSponsored}
{US Department of Homeland Security, Cybersecurity \& Infrastructure Security Agency.}
\newblock People's republic of china state-sponsored cyber actor living off the land to evade detection.
\newblock Online: \url{https://www.cisa.gov/news-events/cybersecurity-advisories/aa23-144a}, 2023.
\newblock Accessed: 16-03-2024.

\bibitem{CISAAdvisoryVoltTyphoon}
{US Department of Homeland Security, Cybersecurity \& Infrastructure Security Agency.}
\newblock Prc state-sponsored actors compromise and maintain persistent access to u.s. critical infrastructure.
\newblock Online: \url{https://www.cisa.gov/news-events/cybersecurity-advisories/aa24-038a?utm_source=CISACyber&utm_medium=post&utm_campaign=VT_020724}, 2024.
\newblock Accessed: 16-03-2024.

\bibitem{10433776}
Muhammad Usama and Muhammad~Naveed Aman.
\newblock Command injection attacks in smart grids: A survey.
\newblock {\em IEEE Open Journal of Industry Applications}, pages 1--11, 2024.

\bibitem{valli2012eavesdropping}
Craig Valli, Andrew Woodward, Clinton Carpene, Peter Hannay, Murray Brand, Reino Karvinen, and Christopher Holme.
\newblock Eavesdropping on the smart grid.
\newblock 2012.

\bibitem{7028837}
Jing Xie, Chen-Ching Liu, Marino Sforna, Martin Bilek, and Radek Hamza.
\newblock Threat assessment and response for physical security of power substations.
\newblock In {\em IEEE PES Innovative Smart Grid Technologies, Europe}, pages 1--6, 2014.

\bibitem{ibm-esri}
Kareem Yusuf.
\newblock How ibm® and esri are working together to map a more sustainable future.
\newblock \url{https://www.ibm.com/blog/how-ibm-and-esri-are-working-together-to-map-a-more-sustainable-future/}, 2023.
\newblock Accessed: 2023-09-09.

\bibitem{industroyer-v2-mandiant}
Daniel~Kapellmann Zafra, Raymond Leong, Chris Sistrunk, Ken Proksa, Corey Hildebrandt, Keith Lunden, and Nathan Brubaker.
\newblock Industroyer.v2: Old malware learns new tricks, 2022.
\newblock Accessed: 29-07-2023.

\bibitem{8228637}
Jiapeng Zhang and Yingfei Dong.
\newblock Cyber attacks on remote relays in smart grid.
\newblock In {\em 2017 IEEE Conference on Communications and Network Security (CNS)}, pages 1--9, 2017.

\bibitem{zheng2013smart}
Jixuan Zheng, David~Wenzhong Gao, and Li~Lin.
\newblock Smart meters in smart grid: An overview.
\newblock In {\em 2013 IEEE green technologies conference (GreenTech)}, pages 57--64. IEEE, 2013.

\end{thebibliography}

%%%%%%%%%%%%%%%%
% Appendices
%%%%%%%%%%%%%%%%
\clearpage
\appendices

\section{Ransomware attacks on power grids}
\label{appendix:annex-A}

\begin{table*}[t]
\centering
\begin{tabular}{llllll}
\toprule
\textbf{Year} & \textbf{Affected Entity} & \textbf{Country} & \textbf{Ransomware Variant} & \textbf{Impact}\\
\midrule
2017 & Iberdrola & Spain & Wannacry & Economic\\

2020 & Reading Municipal Light Department & USA (Massachusetts) & Undisclosed & Economic\\

2020 & LTI Power Systems & USA (Ohio) & Undisclosed & Data Leakage\\

2020 & EDP & Portugal & Ragnar Locker & Economic \& Data Leakage \\

2020 & Northwest Territories Power Corporation & Canada & NetWalker & Operability (Web \& E-Mail)\\

2020 & Elexon & UK & Revil/Sodinokibi & Economic \& Data Leakage\\

2020 & Electricity Generating Authority of Thailand & Thailand & Maze & Data Leakage\\

2020 & Enel Edesur S.A. & Argentina & Snake/EKANS & Economic \& Operability\\

2020 & K-Electric & Pakistan & NetWalker & Economic \& Data Leakage\\

2020 & Enel Group & Global & NetWalker & Economic \& Data Leakage\\

2021 & Carnegie Clean Energy & Belgium & Avaddon & Undisclosed\\

2021 & Centrais Eletricas Brasileiras (Eletrobras) & Brazil & Undisclosed & Operability\\

2021 & Companhia Paranaense de Energia (Copel) & Brazil & Darkside & Data Leakage\\

2021 & Wiregrass Electric Cooperative & USA, Alabama & Undisclosed & Operability (Website)\\

2021 & Delta-Montrose Electric Association (DMEA) & USA, Colorado & Undisclosed & Operability \& Data Loss\\

2021 & CS Energy & Australia, Queensland & Conti  & Operability\\

2022 & ESKOM Hld SOC Ltd. & South Africa & Everest & Economic \& Data Leakage\\

2022 & State Electric Company Limited (STELCO) & Canada & Undisclosed & Operability\\

2022 & Nordex & Global & Conti & Operability\\

2022 & Montenegro government and CI & Montenegro & Cuba & Economic \& Data Leakage\\

2022 & Gestore dei Servizi Energetici SpA (GSE) & Italy & BlackCat/ALPHV & Operability \& Data Leakage \\

2022 & Tata Power & India & Hive & Economic\\

2022 & Empresas Publicas de Medellin (EPM) & Colombia & BlackCat/ALPHV & Operability \& Data Leakage\\

2022 & Entrust Energy & USA, Texas & NetWalker & Economic \\
\bottomrule
\end{tabular}
\caption{Ransomware attacks with impact in the power smart grid}
\label{tab:ransomware-attacks}
\end{table*}

Table \ref{tab:ransomware-attacks} lists known ransomware incidents that impacted electric utilities or the electric grid, obtained from the CIRA dataset~\cite{rege2023critical}. This data is based on news articles or reports that describe the existence of these ransomware cyberattacks, and there might have been more ransomware cyberattacks against the electricity sector. 
We observe an increase in the number of ransomware attacks against the electricity industry in recent years. This can be mainly explained by the fact that ransomware attacks have become massively popular during the in the last few years, especially since the COVID-19 pandemic, and the electricity sector consequently also reflects this increase. 
%It is also important to mention that the dataset that is being used as the main source begins to record data since the year 2017 and is updated until mid 2022~\cite{rege2023critical}. Also the dataset is based on news articles or reports that describe the existence of these ransomware cyberattacks, but it is certain that there have been more ransomware cyberattacks against the electricity sector that the ones recorded in the Table \ref{tab:ransomware-attacks}.
Most of the ransomware cyberattacks have caused economic impact or data leakage, yet only a minority have caused operational damage. In the cases where the operability is affected, it is mainly by bringing down the web or e-mail services of these companies.

\section{Vulnerability analysis}
\label{appendix:annex-B}

In this paper, we modeled adversaries by considering their motivation, goals, knowledge, and capabilities. However, establishing the feasibility of these capabilities has proven to be a intricate task. To avoid merely speculating on potentially unrealistic capabilities, we conducted a study that provides evidence about the practicality of the described threats. This is supported by the identification of vulnerabilities that if properly exploited, give an attacker the mentioned capabilities, validating therefore the realism of our modeled scenario.

This study has been conducted in three phases: selection of devices to be examined, collection of OSINT information about the devices, and analysis of the identified vulnerabilities.

\subsection{Device selection}

Firstly, we have identified a series of devices present in the electrical network that have been or may be targets of attacks by adversaries with the objectives, knowledge, and capabilities we have modeled. For this selection, we have relied on the study of the attack surface carried out in Section~\ref{subsec:Targets}, especially in the operational domain. Based on this, the selected systems and devices are:

\begin{itemize}
\item{Supervisory Control And Data Acquisition (SCADA)}
\item{Geographic Information Systems (GIS)}
\item{Advanced Metering Infrastructure (AMI)}
\item{Supervisory Control And Data Acquisition (SCADA)}
\item{Remote Terminal Unit (RTU)}
\item{Programmable Logic Controller (PLC)}
\end{itemize}

\subsection{OSINT collection methodology}

Once we have identified the main devices to be studied, the next steps is to collect OSINT information for product and manufacturers of these devices and their associated vulnerabilities. For this collection, we leverage two popular catalogs from the NVD, i.e., Common Platform Enumeration (CPE) and Common Vulnerability Exposure (CVE). 

\subsubsection{Products}

To search for specific products, we rely on the CPE system, which is a structured naming scheme for information technology systems, software, and packages, providing a description format for binding text and tests to a name~\cite{nist-cpe}. A key challenge in this step is to identify which are the right IT systems to look for, i.e. those that are used in Smart Grids. However, some of them are not exclusive, e.g. SCADA systems are used in the scope of various ICS. Also, the high market diversity leads to a high amount of products that might be used for Smart Grids, which hardens the analysis. Accordingly, we focus our research on four companies that, according to industry reports~\cite{emergen-research-companies,energydigital-research-companies}, have a higher revenue in the market of Smart Grids by September 2023. These companies are IBM, Cisco, Siemens and ABB. 

\begin{figure}[h]
    \centering
    \includegraphics[width=1\linewidth]{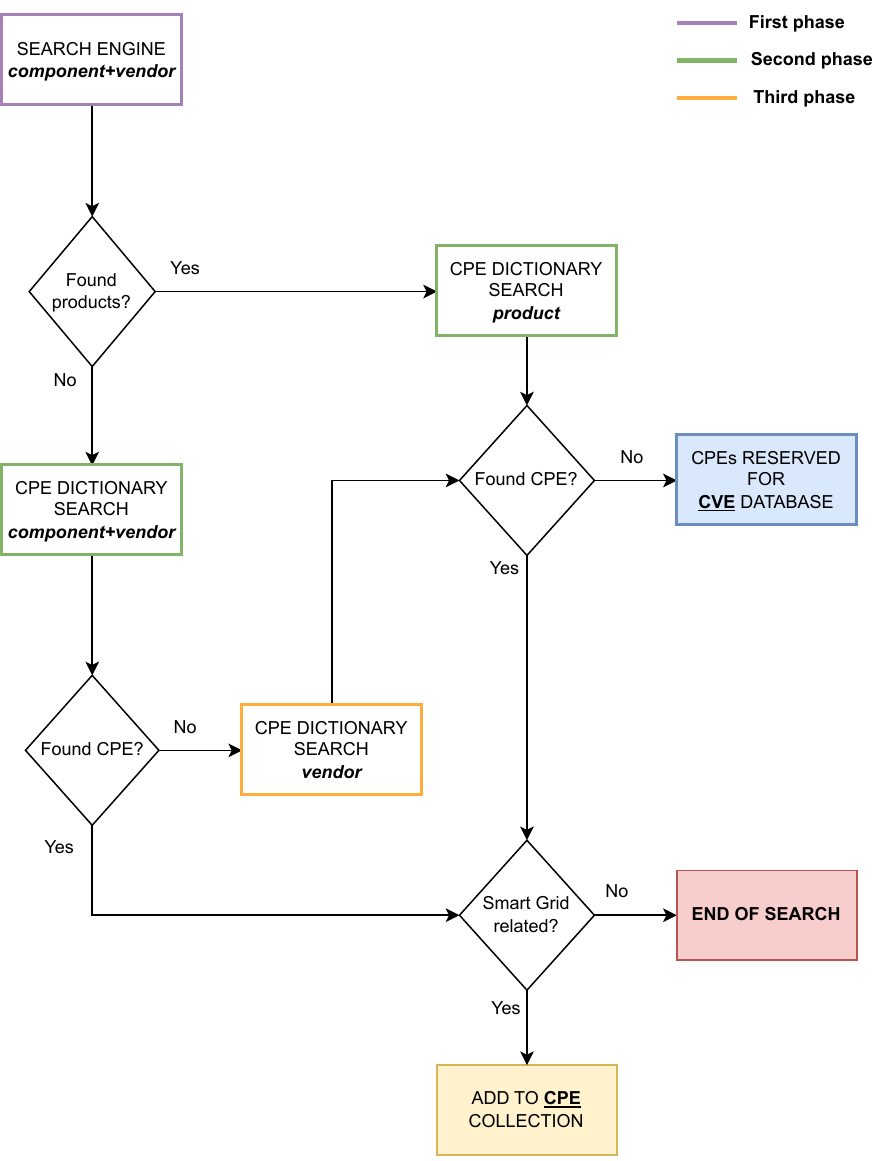}
    \caption{Flowchart of the methodology to collect CPEs}
    \label{fig:flow-method}
\end{figure}

Figure~\ref{fig:flow-method} shows the methodology employed to select the CPEs related to the selected devices. As the flowchart presents, the \textbf{first phase} consisted on introducing the keywords \textit{component} + \textit{vendor} in a web search engine. Using this approach we found a new vendor (Esri), which was not initially identified. We noted that this company has commercial collaboration with Siemens and IBM for manufacturing GIS products, and thus we include it in our analysis~\cite{siemens-esri} \cite{ibm-esri}. 
    
This search allowed to find a catalog of products related to smart grids offered by Siemens \cite{siemens-catalog} providing names for specific products. This allowed us to make specific searches in the CPE Dictionary, as we explain below. Similarly, some of ABB’s specific product names were identified in this stage. However, this search did not give any results about the other two vendors (Cisco and IBM). Although we collected general information about their presence in the smart grid market, it did not mention specific products they may offer for these systems. 

The \textbf{second phase} used the CPE Dictionary. Similar to the previous phase, it uses keywords to find relevant CPEs. In addition to the pair of keywords used in the previous search (\textit{component} + \textit{vendor}), we introduced a new keyword: \textit{product}, to include the specific products found in the first phase.

Finally, we conducted a \textbf{third and final phase} where we searched for products related to \textit{Esri} GIS products, obtaining various CPEs of products related to Smart Grids. In this phase, we made an attempt to look for CPEs related to IBM and Cisco. This search, however, resulted in a huge amount of CPEs (more than 40k results), due to the prevalence of these vendors in various IT markets. Since we could not collect specific information for products being specifically used in Smart Grids, we decided to leave out these two vendors from the study. This way, we choose to not err on the side of having False Positives (i.e., not including devices that are not being deployed in Smart Grids) in our study.
%\sergio{Ref to limitations?}
    
To end this process, all the resulting CPEs were manually verified to confirm that they were used in smart grids, adding these to the collection of CPEs.
For those products that were not found on the CPE dictionary, they were reserved to later search them in the CVE (“Common Vulnerabilities and Exposures”) database.

\subsubsection{CVE search}

\begin{figure}[h]
    \centering
    \includegraphics[width=1\linewidth]{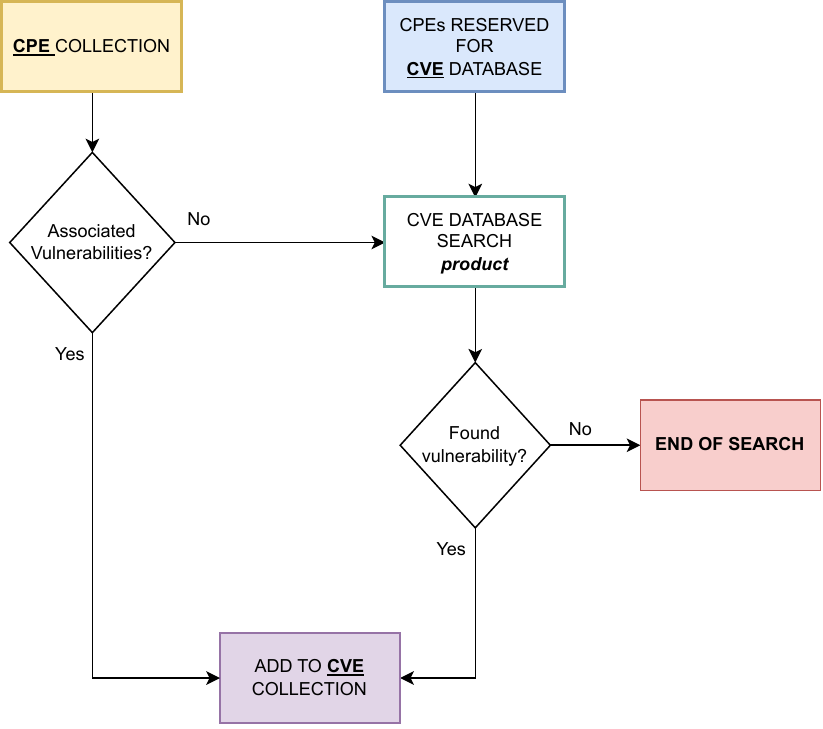}
    \caption{Flowchart of the methodology to gather CVEs}
    \label{fig:flow-cve}
\end{figure}

Figure \ref{fig:flow-cve} depicts the process to find the vulnerabilities by looking for CVEs associated for each of the CPEs collected previously. The set CPEs contain a direct reference to their associated CVEs, which allowed to conduct the query in an automated way. Still, during this process we also searched on the CVE Database for products that did not have an associated CPE, leading to new vulnerabilities on a product from SIEMENS without an associated CPE.

\vspace{0.4cm}

% \subsubsection{Limitations} 
% \sergio{Summarize and explain that we err on the side of not having false positives}

\subsection{Analysis}

%%%%%% CPE Analysis %%%%%%%%

Our collection process resulted on a total of 164 different CPEs, which are then grouped since various of them are different versions for the same product. Accordingly, we finally obtain a set of 50 groups of CPEs (i.e., 50 different products). Table \ref{tab:cpe_group} summarizes the products and CPEs found for each component, and the number of CVEs associated to each of them. It is important to point out that the total of number of CVEs pictured is not the direct sum of all the CVEs, since some CPEs share the same associated CVEs. The group `PLC' refers to Programmable Logic Controllers (PLC) used in Smart Grids. These are general purpose devices that allow to monitor and control ICS, e.g., collecting information or delivering instructions. They can be categorized in any of the other components, and thus are presented in a separate group for clarity.
%The characteristics of this collection are presented in this subsection

\begin{table}[h]
    \centering
    % \resizebox{\columnwidth}{!}{
    \begin{tabular}{l|c|c|c}
    Component &  \#Products & \#CPEs & \#CVEs \\
    \hline
    Remote Terminal Unit (RTU) & 7 & 13 & 11\\
    SCADA & 7 & 8 & 40 \\
    Geographic Info. Sys. (GIS) & 16 & 86 & 83 \\ 
    Power Automation Sys. (PAS) & 9 & 15 & 36 \\
    Advance Meter. Infras. (AMI) & 4 + 1* & 14 + 4* & 17 \\
    Demand Resp. Sys. (DRS) & 1 & 4 & 2 \\
    Program. Logic Contr. (PLC)  & 5 & 20 & 14 \\
    \hline
    \textbf{Total} & \textbf{50} & \textbf{164} & \textbf{203} \\
    \hline 
    \end{tabular}
    % }
    \caption{Overview CPE collection *CPEs that are also present in DRS group}
    \label{tab:cpe_group}
\end{table}

Table~\ref{tab:cpe_vendor} shows the amount of CPEs grouped by vendor, including the type of component. It can be observed that the majority of CPEs belong to the vendor Siemens, which manufactures products of all categories. Additionally, only two CPEs belong to the vendor ABB, corresponding to a PAS and an AMI. Finally, as discussed earlier and as it can be observed in both tables, the vendor `Esri' is exclusively dedicated to the production of GIS products. 

\begin{table}[h]
    \centering
    \begin{tabular}{c|c|c}
    Vendor & Number CPEs & Components present  \\
    \hline
    Siemens & 32 & All \\
    Esri & 16 & 'GIS' \\
    ABB & 2 & 'PAS' and 'AMI' \\ 
    \hline 
    \end{tabular}
    \caption{Distribution of CPEs by vendor}
    \label{tab:cpe_vendor}
\end{table}

%These are some of the characteristics of the products we collected to gather their associated CVEs. Some other features include the names of the products, the groups of CPEs that they conform or external links to their vendor's website, and can be seen in the CPE collection \cite{cpe-collection}.

%%%%%% CVE Analysis %%%%%%%%
Once we have collected the CPEs, we conduct a qualitative overview of the associated vulnerabilities. To this end, we rely on a third metric from the NVD, i.e., the  \textbf{``Common Vulnerability Scoring System'' (CVSS)}. For each vulnerability, this metric characterizes the exploitation capabilities (e.g., whether physical access is required), and the impact in terms of confidentiality, integrity, and availability losses to the affected systems.
We note that this metric is inconsistent and leaves room for ambiguities~\cite{spring2021time}. Indeed, 32 of the collected CVEs contain two different scores, i.e., one provided by the NIST, and another one provided by the vendor. For instance, in \textit{CVE-2022-30694}, NIST set the score as 3,5 indicating a low criticality where Siemens rated it as 6,5 designating a medium criticality. In this case, not only the score changed, but also the assigned criticality level and impact on the CIA Triad. 
In our study, for inconsistent cases, we use the score provided by the vendor, as their role as manufacturers could allow them to have a better understanding on the potential impact of a vulnerability over the NIST.

We analyze the total of 203 CVE collected. We next provide a general overview of the main characteristics of said vulnerabilities. %We provide the entire list of CVEs and CPEs analysed in an online appendix.\footnote{\url{https://anonymous.4open.science/r/adversarial-model-smart-grids-CB2C}}

\begin{figure}[h]
    \centering
    \includegraphics[width=1\linewidth]{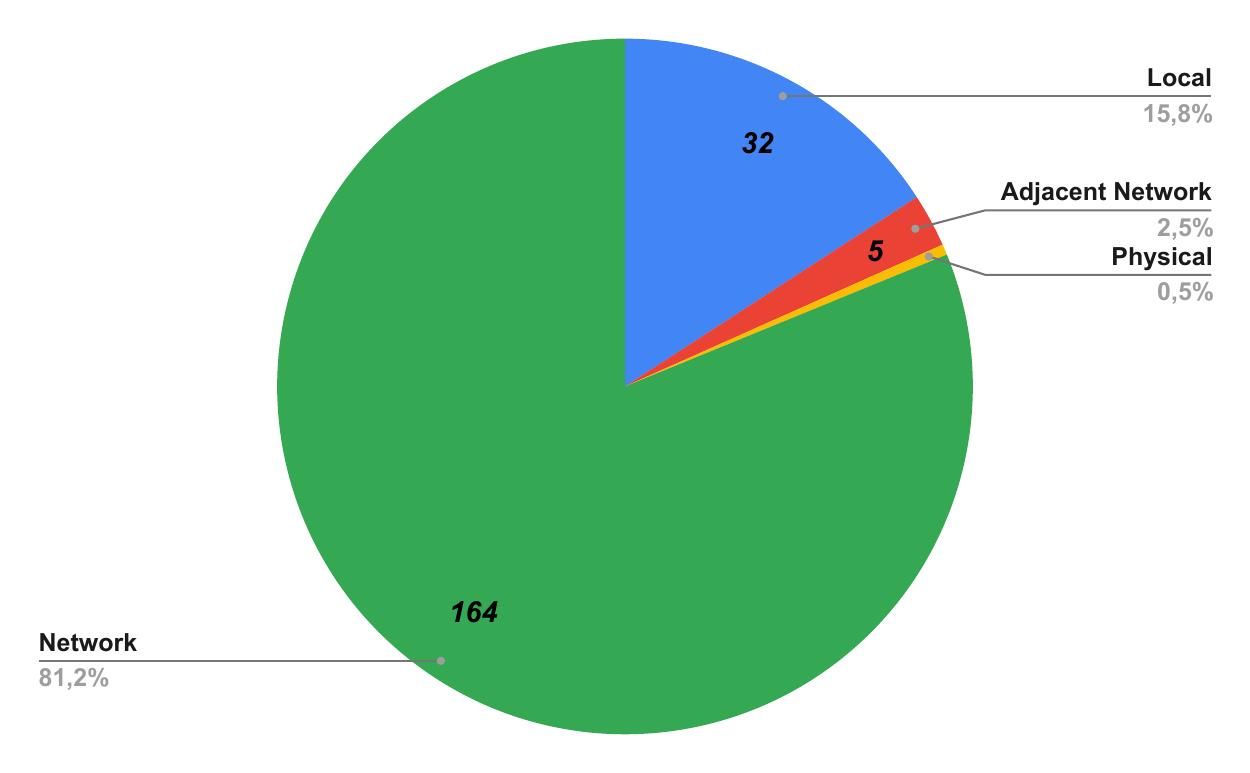}
    \caption{Access vector count of CVE collection}
    \label{fig:access-vector}
\end{figure}

We first analyze the \textbf{access vector}, which indicates the context in which the vulnerability can be exploited. As depicted in Figure~\ref{fig:access-vector}, a vast majority of the vulnerabilities (81.2\%) can be exploited remotely (network access). While this risk can be mitigated by establishing proper perimetral cyber-defenses (e.g., firewalls), a wrong configuration or vulnerability in these defenses could allow an attacker to exploit the device from an external location, posing a risk to the operation of the grid.

\begin{figure}[h]
    \centering
    \includegraphics[width=1\linewidth]{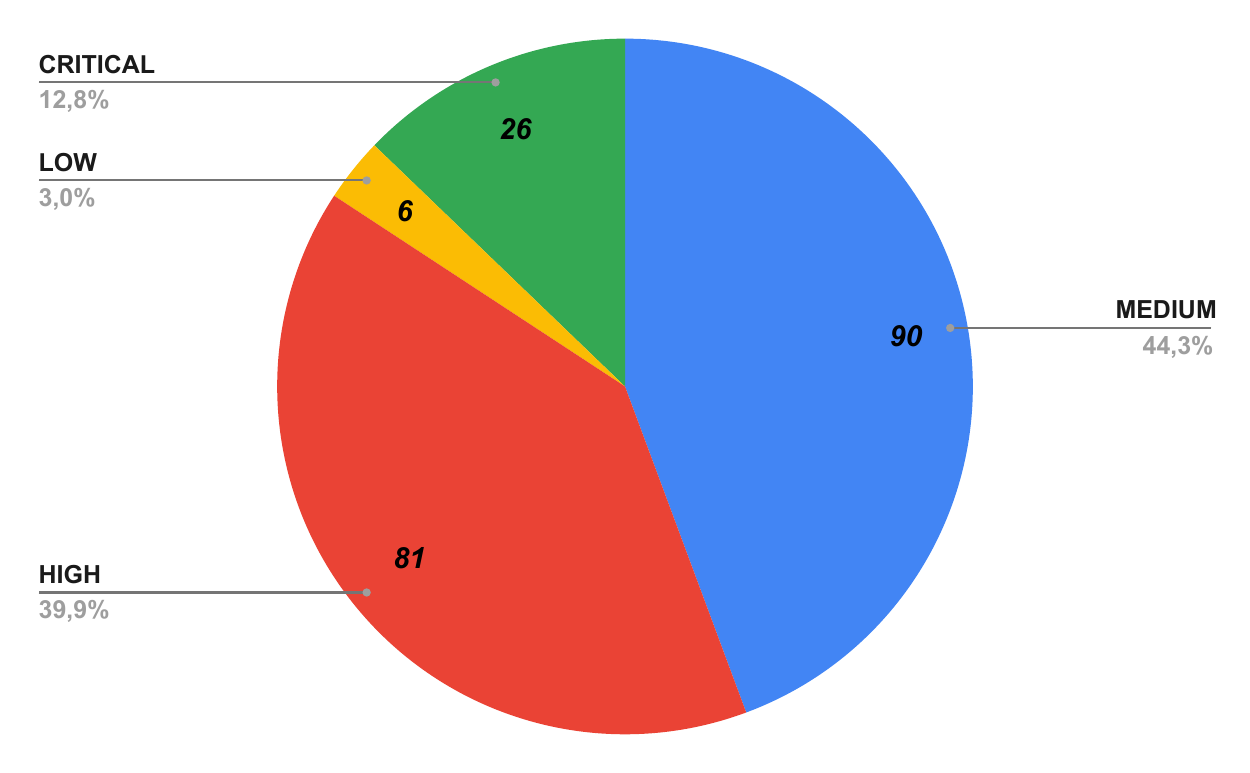}
    \caption{Severity count of CVE collection}
    \label{fig:severity}
\end{figure}

The CVSS assigns a score for the \textbf{severity} of each CVE, based on the impact and exploitability sub scores, which ranges from None to Critical~\cite{nist-score-severity}.  While \textit{Medium} and \textit{Low} scores have lower possibility of being exploited, or their impact is lower, vulnerabilities marked a \textit{High} or \textit{Critical} have a higher risk of being exploited and with more severe consequences. Figure~\ref{fig:severity} overviews the severity of the CVEs identified in this study. Over half of the vulnerabilities have a severity which is high or critical. 

The CVSS base score contains a the feature named ``User interaction'', which can be either `required' if some sort of user activity is required to trigger the exploitation (which can be enforced by means of Social Engineering techniques, e.g., to download a malicious file), or `none' if no user interaction is required, in which case it increases the severity of the vulnerability.
In our case, 123 CVEs ($\approx$ 60\%) did not require user interaction, resulting in higher severity. Among the remaining CVEs, 66 required of user interaction to be exploited while 14 did not have this feature because they were only available in CVSS score version 2, which does not include the user interaction metric. 
Other CVSS features have a similar impact in the overall score. However, it is important to know that that low severity vulnerabilities are still exploitable, as sophisticated attackers could use them to carry out complex attacks due to the interdependencies of the different devices for the proper operation of the grid (see \S\ref{sec:attacks}). 
\\ 

\begin{table}[h]
    \centering
    \begin{tabular} {c|c|c|c}
    Impact & Confidentiality & Integrity & Availability \\
    \hline
    None & 50 - (24,6\%) &  63 - (31,2\%) & 94 - (46,3\%) \\
    Low & 62 - (30,5\%)  & 64 - (31,7\%) & 11 - (5,4\%) \\
    High & 91 - (44,8\%)  & 75 - (37,1\%) & 98 - (48,3\%) \\
    \hline
    \end{tabular}
    \caption{CIA Triad impact of CVE collection}
    \label{tab:cia-impact}
\end{table}

Table~\ref{tab:cia-impact} presents the impact that the vulnerabilities have on the three security domains: confidentiality, integrity and availability. 
%According to the NIST, confidentiality and integrity impact refer to the effects on the information resources of a product, while availability impact refers to the impact on the product itself.  
As shown in the table, more than half of the vulnerabilities would have an impact on the \textbf{availability} of a product, which could cause it to stop its function. This is particularly worrying due to interdependencies of the devices in the different operational domains. Therefore, if a product (or group of products) were to stop functioning and providing service, it would have an impact on the rest of the domains of the grid. It is important to point out that smart grids are critical infrastructures. Therefore, any  impact on the availability of a product is critical as it could disrupt the correct functioning of the grid.

In terms of \textbf{integrity}, over two thirds of the vulnerabilities have impact on the integrity of the information provided by the products, with 37\% having a high impact. These numbers are concerning as smart grids rely on the data recorded by different products across the grid to determine the best electricity distribution, and tampering with these data could provoke important financial or operational damage. Indeed, a compromise on the integrity of the data could cause the energy flow to be below or over its required level. This could result in insufficient energy for consumers or product damage, and overall, a malfunctioning of the grid. 

And in regard to the \textbf{confidentiality}, over than 75\% of the CVEs have impact on confidentiality where almost 45\% of the total are classified as high impact. This raises privacy concerns because as explained earlier in the paper, smart grids receive data from the consumer domain. Consequently, the information of the consumers is also at risk. Additionally, leaked information, such as the consumed or generated electricity, could be used to exploit vulnerabilities with integrity impact, providing better understanding of the flow of electricity across the network.
%
%\section{Gory Details}

\end{document}